\documentclass[letterpaper,11pt]{article}
\usepackage{style}
\usepackage{shortcuts}
\usepackage{hyperref}
\usepackage{algorithm2e}
\usepackage{soul}

\newcommand{\cabs}[1]{\abs{#1}_{\pm\Delta}}

\title{Real-to-Integer Reduction for All-Pairs Shortest Paths}
\author{
  Nick Fischer  \\ \parbox[t]{4cm}{\small\centering Max-Planck Institute\\for Informatics} \and
  Adam Polak    \\ \parbox[t]{4cm}{\small\centering Bocconi University}                   \and
  Jonas Schmidt \\ \parbox[t]{4cm}{\small\centering Bocconi University}
}
\date{}

\begin{document}

\maketitle

\begin{abstract}
\noindent
Efficient graph algorithms often use techniques that are specific to integer-valued inputs and do not work for real-valued versions of the problems. Is it a limitation of current techniques, or are the real-valued variants inherently harder? One way to study this question is through real-to-integer reductions. We give (randomized, combinatorial) subcubic real-to-integer reductions for two central graph problems in fine-grained complexity: All-Pairs Shortest Paths and Exact Triangle. That is, we show that if the integer variants of these problems can be solved in truly subcubic time, then the real variants can also be solved in truly subcubic time.

Our work extends and combines techniques from three previous works: (1) our real-to-integer reduction for the 3SUM problem (Fischer, Polak, Schmidt, FOCS 2026), which itself combines the Frank--Tardos theorem from optimization with Freiman's theorem from additive combinatorics; (2) universe reduction for APSP (Fischer, STOC 2026), which we generalize to reals and use to reduce the doubling of the set of edge weights; and (3) the ``Fredman's trick meets dominance product'' technique (Chan, Vassilevska Williams, Xu, STOC 2023), which is needed to adapt the prior real-to-integer framework from equality constraints to inequalities.

As a corollary, our reduction combined with Yuval's fast Integer APSP algorithm in the real RAM model with multiplication (IPL 1976) shows that the Real APSP Hypothesis is false in this model, which also follows from the very recent breakthrough by Alman and Vassilevska Williams using a very different approach.

\vskip 1em

\noindent\textbf{AI methodology.}\quad
We obtained all the results in this paper without any use of AI tools. We used OpenAI's GPT and Google's Deep Review only for final proofreading of the paper.
\end{abstract}

\thispagestyle{empty}
\setcounter{page}{0}
\clearpage

\tableofcontents
\thispagestyle{empty}
\setcounter{page}{0}
\clearpage

\section{Introduction} \label{sec:intro}

When designing fast algorithms for (weighted) graph problems, we have available many techniques that work only for integer edge weights, e.g., scaling~\cite{Gabow85}, bit-based priority queues~\cite{Boas75,AhujaMOT90}, encoding weights as polynomial exponents~\cite{AlonGM97}, or iterative methods whose convergence hinges upon integrality~\cite{Ford56}. Many graph problems seem to become substantially harder when we allow for real-valued weights. A classic example of this phenomenon is the negative-weight Single-Source Shortest Paths (SSSP) problem: for integer edge weights, an improvement over the textbook $O(nm)$-time\footnote{Throughout the paper $n$ denotes the number of nodes and $m$ the number of edges of the input graph.} Bellman--Ford algorithm was found already in the 1980s~\cite{Gabow85}, and now we know a near-linear time algorithm~\cite{BernsteinNW25}, whereas for real edge weights the first improvement over Bellman--Ford came only in 2024~\cite{Fineman24} and the existence of an $\tilde O(n+m)$-time\footnote{We use the $\tilde O$ notation to suppress polylogarithmic factors, i.e., $\tilde O(T) = T \log ^{O(1)} T$.} algorithm is still open. The maximum flow problem has a similar history~\cite{GoldbergR98,ChenKLPGS25}, but there only polylogarithmic improvements over the $O(nm)$ running time are known for real-valued capacities~\cite{CheriyanHM90,Orlin13}.

The main focus of this paper is another fundamental graph problem, the All-Pairs Shortest Paths (APSP) problem. Given a directed edge-weighted graph $G=(V,E,w)$, with no negative cycles, the APSP problem asks to compute the length of a shortest path from $u$ to $v$ for every pair of vertices $(u,v) \in V \times V$.
APSP is one of the central problems in fine-grained complexity~\cite{VassilevskaW18,Vassilevska18}.
The textbook Floyd--Warshall algorithm~\cite{Floyd62a,Warshall62,Roy59} solves it in time $O(n^3)$, and for a long time, until the very recent $O(n^{2.9995})$-time breakthrough algorithm~\cite{AlmanV26}, only a subpolynomial $2^{O(\sqrt{\log n})}$ improvement was known~\cite{Williams18}. This state of affairs led people to study APSP through the lens of \emph{subcubic} reductions---i.e., reductions that show that a truly subcubic algorithm for one problem yields a truly subcubic algorithm for another problem---and many problems are now known to be equivalent to APSP under such reductions~\cite{VassilevskaW18,TamakiT98,BackursDT16,AbboudGW23,BringmannGMW18,NoglerPSVWXY25}.

For APSP, the difference between integer- and real-valued inputs is less immediately visible, because the best known algorithms (both currently and historically) have the same running times. However, the techniques differ substantially. For instance, the paper of Williams~\cite{Williams18} begins with ``a succinct exposition of a good algorithm for APSP, at least in the case
of reasonable-sized [integer] weights'', which is followed by a much longer description of the full algorithm for real-valued inputs. Also, the recent breakthrough truly subcubic algorithm for integer inputs is deterministic while for reals it is randomized~\cite{AlmanV26}.

This phenomenon, for APSP and other problems, begs to ask a question:
\begin{center}
    \itshape
    Are integer-only tricks a necessity or just a convenience?
\end{center}

We remark that this distinction is related, but not equivalent, to the distinction between algorithms whose running time is weakly-polynomial (i.e., it depends polynomially on the bit-length of input numbers) and strongly-polynomial (i.e., it does not depend on the size of input numbers). Often, a strongly-polynomial time algorithm for integers would also work for reals, but this is not always the case (e.g., \cite{BaranDP08}).

\subsection{Our Results}

We address the above question in the case of APSP by providing a real-to-integer subcubic reduction.

\begin{theorem}[Real-to-Integer Reduction for APSP]\label{thm:apsp}
    If Integer APSP, with edge weights from $\{-u,\dots,u\}$, can be solved in time $O(n^{3-\epsilon}(\log u)^{O(1)})$ for some constant $\epsilon > 0$, then Real APSP can be solved in time $O(n^{3-\Omega(\epsilon^3)})$, by a randomized algorithm correct with high probability.
\end{theorem}

\Cref{thm:apsp} shows that, for APSP, as long as the goal is a truly subcubic running time, anything that can be achieved with integer-only tricks automatically applies to real-valued inputs as well.

Naturally, the implication from \Cref{thm:apsp} is no longer as interesting as it was before the recent work of Alman and Vassilevska Williams~\cite{AlmanV26}, which shows that both sides of this implication are \emph{unconditionally} true. Nonetheless, we believe that the extensive techniques behind our proof of \Cref{thm:apsp} bring us closer to understanding the driving question of this paper about the limitations of integer-specific tricks in algorithm design.

Alman and Vassilevska Williams~\cite{AlmanV26} propose that now, when we know that the original APSP Hypothesis is false, ``one could refocus the hypotheses to be about combinatorial algorithms.'' We remark that our reduction from \Cref{thm:apsp} is combinatorial (i.e., it does not use algebraic Strassenesque techniques common to fast matrix multiplication algorithms). Hence, we also establish that the ``Combinatorial Real APSP Hypothesis'' (asserting that no ``combinatorial'' algorithm for Real APSP runs in truly subcubic time in the real RAM model) implies the ``Combinatorial Integer APSP Hypothesis'' (asserting that no ``combinatorial'' algorithm for Integer APSP runs in truly subcubic time in the Word RAM model).

\paragraph{Subcubic Real APSP}

Curiously, in the real RAM model with multiplications (but even without flooring), there is an algorithm that solves Integer APSP in time $n^{\omega + o(1)}$~\cite{Yuval76,ChanVX22}, where $\omega < 3$ denotes the matrix multiplication exponent. In combination with our real-to-integer reduction from \Cref{thm:apsp}, we hence get an algorithm solving Real APSP in truly subcubic time in this model. 

\begin{corollary}[Real APSP Hypothesis Is False in Real RAM Model with Multiplication]
    There exists a constant $\epsilon > 0$, such that Real APSP can be solved in time $O(n^{3-\epsilon})$ in the real RAM model with multiplications, by a randomized algorithm correct with high probability.
\end{corollary}

We note that the recent independent work~\cite{AlmanV26}, using a very different algorithm, establishes essentially the same theorem but in a more reasonable real RAM model with only comparison and addition (see Section~\ref{sec:model} for a discussion of computational models for reals).

\paragraph{Exact Triangle.}

APSP is known to be equivalent under subcubic reductions to the so-called Negative Triangle problem~\cite{VassilevskaW18}, which asks, given an edge-weighted graph, if there exists a triangle (a cycle of length $3$) with a negative total weight. A related problem, Exact Triangle is another central problem in fine-grained complexity. It asks if there is a triangle with total weight equal to zero. Negative Triangle (and hence also APSP) reduces to Exact Triangle~\cite{WilliamsW13}, but there is no known reduction in the opposite direction; Exact Triangle seems to be a genuinely harder problem. Also for this problem we give a real-to-integer reduction:

\begin{theorem}[Real-to-Integer Reduction for Exact Triangle]\label{thm:exact-triangle}
    If Integer Exact Triangle, with edge weights from $\{-u,\dots,u\}$, can be solved in time $O(n^{3-\epsilon}(\log u)^{O(1)})$ for some constant $\epsilon > 0$, then Real Exact Triangle can be solved in time $O(n^{3-\Omega(\epsilon^2)})$, by a randomized algorithm correct with high probability.
\end{theorem}

It is not a coincidence that we also prove this result; there is a technical reason why we get it along the way to \Cref{thm:apsp}. We elaborate on it in \Cref{sec:overview}, but let us briefly explain this reason now: Our starting point is our previous real-to-integer reduction for the 3SUM problem~\cite{real3sum}, the first such self-reduction in fine-grained complexity. To extend our 3SUM framework to APSP, we need to account for the two main differences between these two problems: first, 3SUM is an arithmetic/convolution problem while APSP is a graph/matrix problem; second, 3SUM is about equality constraints and APSP is about inequalities. The Exact Triangle problem is a graph/matrix problem about equality constraints, so it is a natural warm-up that lets us understand how to deal with the first difference without yet having to worry about the second one.

\subsection{Computational Model}
\label{sec:model}

When working with real-valued inputs, one has to be careful about the choice of computational model. Obviously one cannot just work with the word RAM, which is the standard choice (for problems with integral inputs) in fine-grained complexity, because it lacks the ability to express real numbers. Its analog for reals, the real RAM model, quickly becomes unreasonably powerful if we allow for arbitrary arithmetic operations on infinite-precision numbers in constant time. For instance, flooring allows solving PSPACE-complete problems in polynomial time~\cite{HartmanisS74,Schonhage79}. Even without flooring, multiplication itself allows solving  APSP with integer edge weights in the matrix multiplication time~\cite{Yuval76}. For our reductions to be meaningful, we want to state them in a model as weak as possible.

All our reductions work in the \emph{comparison-addition} model, which is the word RAM model augmented with only two real-valued operations: comparisons and additions. Actually, they even work in the so-called \emph{realistic} comparison-addition model \cite[Section~1.1]{real3sum}, where the cost of evaluating a comparison of the form $\sum_i n_i\alpha_i \geq 0$, where $n_i$'s are arbitrary non-zero integer coefficients and $\alpha_i$'s are input reals, is $\smash{\big(\sum_i(1+\log|n_i|)\big)^{O(1)}}$, and real addition can be performed only symbolically. A benefit of working with this model is that any algorithm for real-valued inputs can be efficiently simulated in the word RAM model on inputs that happen to be integral. In other words, trivial integer-to-real reductions do work with this model choice, and, combined with our real-to-integer reductions, they give fine-grained, subcubic \emph{equivalences} of the real and integer variants of APSP and Exact Triangle. See \cite[Section~1.1]{real3sum} and \cite[Appendix A in the full version]{ChanVX22} for a deeper discussion of computational models for reals.

\subsection{Open Problems}

A natural direction for future research is to make our real-to-integer reduction more efficient. Ideally, one would like an $O(n^{3-\epsilon})$-time algorithm for Integer APSP to imply an $\tilde O(n^{3-\epsilon})$-time algorithm for Real APSP, but already improving the savings from $\Omega(\epsilon^3)$ to $\Omega(\epsilon)$ would be interesting. Let us mention that, if one does not insist on the reduction being combinatorial, then it is already possible to improve $\Omega(\epsilon^3)$ to $\Omega(\epsilon^2)$ by using Matou{\v s}ek's fast algebraic dominance product algorithm~\cite{Matousek91} in the proof of \Cref{lem:detect-very-negative}.

In general, there are many fundamental problems for which the fastest known algorithms for reals are slower than what we can achieve with integer-specific techniques, e.g., linear programming~\cite{Dantzig65,Khachiyan79} or maximum flow~\cite{Orlin13,ChenKLPGS25}. Can we replace these techniques with more general ones, which also handle reals, possibly via real-to-integer reductions?

\section{Technical Overview} \label{sec:overview}

In this section we give a high-level overview of our results. They can be seen as a conceptual and technical extension of the real-to-integer framework from our earlier paper on the 3SUM problem~\cite{real3sum}. This extension concerns two orthogonal directions: from the arithmetic setting to graphs, and from equality constraints ($a + b = c$) to inequality constraints ($a + b \leq c$). The transition from arithmetic to graph problems is technically tedious, but the transition from equalities to inequalities is the actual challenge. There are two concrete obstacles, namely two integer-specific tricks that need to be replaced with alternatives for reals (to be discussed soon), and overcoming them
is the key technical contribution of our paper.

\subsection{Real-to-Integer Framework}
\label{sec:framework}
Before we can get to these new obstacles, it is instructive to understand the key ideas behind our earlier real-to-integer framework. For simplicity, focus on the 3SUM problem, which asks to detect whether a given set $A$ contains a solution to the equation $a + b + c = 0$. In~\cite{real3sum}, we show that this problem can be reduced from real-valued inputs to integer-valued inputs, in terms of subquadratic-time algorithms.

The starting point of that work is the Frank--Tardos theorem from optimization, which can be stated as follows:

\begin{theorem}[Frank--Tardos~\cite{FrankT87}] \label{thm:frank-tardos}
There is an algorithm that, given reals $\alpha \in \Real^d$, runs in polynomial time $d^{O(1)}$ and computes integers $v \in \Int^d$ such that
\begin{enumerate}[label=(\roman*)]
    \item $\norm{v}_\infty \leq 2^{O(d^3)}$, and
    \item $\alpha^T x \geq 0$ if and only if $v^T x \geq 0$, for all $x \in \set{-1, 0, 1}^d$.
\end{enumerate}
\end{theorem}

Suppose we apply this theorem to the reals in our input set $A = \set{a_1, \dots, a_n}$. Then the theorem supplies integers $v_1, \dots, v_n$ that can be directly used as substitutes for these reals; indeed, there is a real solution $a_i + a_j + a_k = 0$ if and only if there is an integer solution $v_i + v_j + v_k = 0$. While this is a fascinating starting point, the problem is that running the Frank--Tardos theorem is too slow: the algorithm runs in polynomial time with some large exponent, and even if this exponent could be improved, it is unconditionally impossible to achieve subquadratic time. (For details, see~\cite{real3sum}.)

Therefore, instead of applying the Frank--Tardos theorem directly,~\cite{real3sum} proceeds differently. Here we describe this approach at a very high level, omitting many technical details without mentioning them. The approach can be summarized in the following steps:
\begin{enumerate}
    \item\emph{Rounding:} Find an appropriate \emph{scale} $\Delta \in \Realnneg$ and round all reals to integer multiples of $\Delta$. Of course, we cannot actually round reals---this operation is not supported in the real RAM model---so we instead work with the usual weights but with $\pm \Delta$-approximate equalities. But, for simplicity, let us pretend in this overview that we can round.
    
    The scale $\Delta$ is chosen in such a way that the number of solutions after the rounding is close to the maximum $n^2$, say $\Theta(n^{2-\epsilon})$.\footnote{\label{foot:conv-3sum}Strictly speaking, after rounding the number of solutions could even be as large as $n^3$. For this reason, in~\cite{real3sum} we work with the \emph{Convolution 3SUM} problem where the goal is to detect a solution $a_i + a_j + a_k = 0$ for which additionally $i + j = k$. This restriction gives a naive upper bound of $n^2$ on the number of solutions. This technicality shall not concern us here.} This has an advantage and a disadvantage: the advantage is that we have introduced \emph{additive structure} by rounding at this coarse-grained scale, and we will soon exploit that structure. The disadvantage is that we have introduced many \emph{false positive} solutions, and in order to distinguish a true solution from these false positives, we are forced to list all rounded solutions.
    \item\emph{Popular/unpopular decomposition:} While the rounding has introduced a lot of additive structure globally, this does not mean that every item is affected by this structure locally. We therefore partition the elements into \emph{popular} and \emph{unpopular} ones, depending on their local additive structure.\footnote{This is an oversimplification. In fact, in~\cite{real3sum} we need to consider a third category, \emph{quasi-popular} items. This is for technical reasons that we ignore here.} Concretely, we say that an element $a_i$ is popular if it participates in many solutions of the form $a_i + a_j = a_k + a_\ell$ (after the rounding).\footnote{Due to the same technical issue discussed in \cref{foot:conv-3sum}, here we only consider the solutions that also satisfy $i + j = k + \ell$, in fact.} The idea behind this decomposition is that the popular entries can be replaced by integers: via some involved additive combinatorics machinery, one can cover the popular entries by the \emph{subset sums} of a set $X$ of size at most~$d = n^{O(\epsilon)}$ (a~so-called \emph{subset sum cover}). Then we can apply the Frank--Tardos theorem to $X$ instead of $A$, which runs in time $n^{O(\epsilon)}$.
    \item\emph{Listing rounded solutions:} We now list all rounded solutions. Of course, replacing only the popular items by integers would not be fully sufficient, as in a solution $a_i + a_j + a_k = 0$ not all items need to be popular. We are left with three cases:
    \begin{itemize}
        \item[3.1.] \emph{All-popular solutions:} First, consider solutions $a_i + a_j + a_k = 0$ where all three items are popular. These can now be listed, by replacing popular reals with their integer representatives, and then using the integer algorithm (by basic tricks from fine-grained complexity, any detection algorithm can be turned into a listing algorithm). To keep this step efficient, we rely crucially on the guarantee from Step~1 that the number of rounded solutions is at most $O(n^{2-\epsilon})$. 
        \item[3.2.]\label{step:partpop} \emph{Partially popular solutions:} Second, consider solutions $a_i + a_j + a_k = 0$ where some item is popular and some other item is unpopular. We cannot detect these solutions by a reduction to integers. Instead, we use a technical argument (a version of the ``Fredman's trick meets dominance product'' technique from~\cite{ChanVX23}), exploiting both that solutions involving popular items can be efficiently listed (due to their structure), and that there are not many solutions involving unpopular items (due to their missing structure).
        \item[3.3.] \emph{All-unpopular solutions:} It remains to list solutions containing three unpopular items. To do so, we simply \emph{remove} all popular items and iterate, starting over in Step 1. This is clearly correct, but not necessarily efficient, as we now have to worry about the number of iterations. In particular, if the number of iterations is $n^\epsilon$ then already listing the all-popular solutions across all iterations becomes too expensive. The insight is that, since there are many rounded solutions, at least $\Omega(n^{2-\epsilon})$, and since every unpopular item participates in few, a large fraction of items must be popular. We thus reduce the size of $A$ considerably with each iteration, leading only to $\tilde O(n^{\epsilon/2})$ iterations in total.  
    \end{itemize}
\end{enumerate}

\subsection{APSP}
Let us finally come back to APSP. A seminal result in fine-grained complexity is that APSP is equivalent (in terms of subcubic algorithms) to the \emph{Negative Triangle} problem, asking whether a given tripartite graph contains a triangle $(i, k, j)$ with a negative total edge weight,
\begin{equation*}
    w(i, k) + w(k, j) + w(i, j) < 0.
\end{equation*}
A closely related problem is \emph{Exact Triangle}, where the goal is instead to decide if there is a triangle with the total weight equal to zero,
\begin{equation*}
    w(i, k) + w(k, j) + w(i, j) = 0.
\end{equation*}
That is, Exact Triangle can be seen as a variant of 3SUM where the items have to be placed on the edges of a triangle, and Negative Triangle can be seen as the variant with ``$<$'' instead of ``$=$''.

An adaptation of the real-to-integer reduction should therefore be implemented in two steps: (1) from 3SUM to Exact Triangle, i.e., from an arithmetic to a graph problem, and (2) from Exact Triangle to Negative Triangle, i.e., from ``$=$'' to ``$<$''. These two steps yield separate challenges. The first one is simpler; at least conceptually, the approach from before can work (though there are technical issues). The second one is much more challenging, and in the following overview we only focus on the two concrete obstacles that need to be overcome.

\paragraph{Problem 1: Additive Hashing.}
We start with the bigger of the two problems. An important aspect of the real-to-integer framework that we have not highlighted before is the following: the integers resulting from the Frank--Tardos theorem are huge! Even when applied in the elaborate way described above, through additive combinatorics and subset sum covers, we call the Frank--Tardos theorem with $d = n^{\epsilon}$ and thus the resulting integers have magnitude $\exp(d^{O(1)}) = \exp(n^{O(\epsilon)})$. In particular, performing any arithmetic operation on these numbers takes time $n^{O(\epsilon)}$. We can certainly not afford to lose this extra factor in the running time, since the saving in the aforementioned reduction is less than $n^{\epsilon/2}$.

For this reason, an important intermediate step, after we have applied Frank--Tardos but before we list integer solutions, is to reduce the size of these integers; in other words, we perform a \emph{universe reduction}. For 3SUM this is simple: all we need is a hash function $h$ that maps arbitrarily large integers to a small universe, say $\set{-n^{O(1)}, \dots, n^{O(1)}}$, and that is \emph{additive} in the sense that $a + b + c = 0$ if and only if $h(a) + h(b) + h(c) = 0$ (with good probability). Many families of additive hash functions are known, e.g., $h(x) = x \bmod p$ for a randomly chosen prime $p$ (and then working with equality modulo $p$).

If we want to apply the same approach for Negative Triangle, we would need a hash function satisfying the analogous property that $a + b + c < 0$ if and only if $h(a) + h(b) + h(c) < 0$ (with good probability). However, such a hash function to a small universe cannot exist!\footnote{To see this, consider the set $A = \set{0} \cup \set{-3^i, 3^i : i \in \{0,\dots,n\}}$ and suppose that there is a hash function $h : A \to \Int$ satisfying this property. We sketch the argument assuming that the property holds deterministically, but it can be extended to the randomized guarantee. Since $0 \in A$, each value $h(a)$ is positive/zero/negative if and only if $a$ is positive/zero/negative. Moreover, we clearly have $-3^i + 3^{i-1} + 3^{i-1} < 0$ for all $i$, and thus $|h(-3^i)| > 2|h(3^{i-1})|$. Similarly, $|h(3^i)| \geq 2|h(-3^{i-1})|$. Iterating this, we must have that $|h(3^n)| \geq 2^n |h(1)|$; i.e., the range of $h$ is exponential in $n$.}  This forces us to take a different route which, we believe, is of independent conceptual interest.

\paragraph{Solution 1: Doubling Reduction}
As we are looking for a universe reduction for Negative Triangle (i.e., APSP) from exponential to polynomial weights, perhaps a natural starting point is Fischer's recent universe reduction from polynomially-bounded to linear weights~\cite{Fischer26}. Let us take a step back to describe some ideas behind that work. Instead of directly using a hash function to reduce the universe size, it suggests a detour in three steps: (i) first, round the integers so that the set of weights $A = \set{ w(i, j) : i, j \in [n]}$ admits some weak structure, (ii) then amplify this weak structure into much stronger structure called \emph{low doubling}, (iii) and only then hash the weights to a small universe. Steps (i) and (iii) are similar to our setup here, but the crucial step is to apply the \emph{doubling reduction} in between. Intuitively, when the weights are structured in this sense we circumvent the counterexamples showing that the desired hash functions cannot exist in general. To define low doubling for an integer set~$A$, let $A + A = \set{a + b : a, b \in A}$ denote its \emph{sumset}. The \emph{doubling} is defined as the ratio $K := |A + A| / |A|$. Fischer's doubling reduction essentially allows for the following trade-off: the weights can be transformed to have arbitrarily small doubling~$K$, at the cost of an overhead of $O(n^3 / K^{\Omega(1)})$ in the running time. In particular, $K$ can be an arbitrarily small polynomial.

Unfortunately, the doubling reduction is long and involved. It consists of five steps, four of which need to be reproved for our setting of approximate reals (recall that we always work with reals represented up to $\pm \Delta$). In \cref{sec:proof-low-rank-to-small-doubling} we work through these four steps, and along the way also reprove a version of the famous Balog--Szemerédi--Gowers (BSG) theorem for approximate reals. But with the doubling reduction in place, we can repair the approach from before as follows. We replace the second step of the real-to-integer framework (Section~\ref{sec:framework}) with the following more elaborate version:
\begin{enumerate}
    \item[2.1.] Decompose the weights into popular and unpopular ones. Our goal in the following steps is to replace the popular ones by \emph{small} integers.
    \item[2.2.] Run the doubling reduction on the popular weights. This allows us to assume that the popular weights have doubling $K = n^\delta$ for some arbitrarily small $\delta$. It still remains to replace them by small integers.
    \item[2.3.] Apply Freiman's theorem from additive combinatorics to obtain a small subset sum cover $X$ of the popular weights. One can achieve that $X$ has size $d = |X| \leq (K \log |A|)^{O(1)} = n^{O(\delta)}$. Then use the Frank--Tardos theorem, applied to $X$, to replace the popular weights by integers of size $\exp(d^{O(1)}) = \exp(n^{O(\delta)})$.
\end{enumerate}
Thus, the resulting integers are still exponentially large, and performing operations on them takes time $n^{O(\delta)}$. However, the new freedom, due to the doubling reduction, is that we can pick $\delta$ arbitrarily small, independent of the other parameters of the reduction. In particular, when setting $\delta$ sufficiently small in terms of $\epsilon$ we can make sure that the total overhead due to the large weights becomes, say, $O(n^{\epsilon/4})$, and so the real-to-integer reduction stays efficient. This resolves the first problem.\footnote{The structure amplification in the doubling reduction even accomplishes slightly more: It allows us to simplify the overall reduction by removing the need for a separate step like Step 3.2 (\Cref{sec:framework}), the partially popular case. These solutions can now also be detected by a reduction to integers.}

There are two remaining technical issues worth commenting on. First, if Step 2.2 runs in time $O(n^3 / K^{\Omega(1)}) = O(n^{3-\Omega(\delta)})$, and we repeat this step over the $\tilde O(n^{\epsilon/2})$ iterations, why does this not cancel all of our improvements, given that $\delta$ is chosen to be much smaller than $\epsilon$? The insight is that every popular edge appears in the doubling reduction in at most one iteration. Thus, analyzing the $O(n^3 / K^{\Omega(1)})$ running time in terms of the number of edges rather than vertices shows that the total time remains subcubic.

Second, how can we implement Freiman's theorem from additive combinatorics? This problem has been considered by Randolph and W{\k{e}}grzycki~\cite{RandolphW24}; however, their algorithm has an FPT-dependence on the doubling $K$ which is prohibitively slow for us. Luckily, as a side product of our earlier work~\cite{real3sum}, we already have a weaker ``Poor Man's'' version of the theorem which can be implemented sufficiently fast. 

\paragraph{Problem 2: Scaling}
The second problem concerns the standard ``scaling trick'' that is used to reduce from Negative Triangle to Exact Triangle, i.e., from ``$<$'' to ``$=$'' constraints. More specifically, this trick allows us to reduce the Negative Triangle problem to few instances of Exact Triangle, where each instance is promised to only contain nonnegative triangles. It essentially works by determining the weight of the most negative triangle bit-by-bit. However, for infinite-precision reals, such a bit trick clearly cannot work. The same problem has occurred in~\cite{ChanVX22}, where the goal was to reduce from APSP via Exact Triangle to the All-Edges Sparse Triangle problem; their solution was to circumvent the detour via Exact Triangle altogether. Unfortunately, this is not an option for us.

Concretely in our setting, the scaling trick is still necessary for many technical arguments under the hood. For instance, the doubling reduction is built on various heavy/light tricks, which reduce \emph{some} weights to have low doubling, but for other weights they simply enumerate all zero-weight triangles involving these weights. Hence, without the scaling trick we could potentially miss some negative triangles.

\paragraph{Solution 2: Fredman's Trick meets Dominance Product}
Our solution is as follows. Suppose that we still follow our framework from before; in each iteration, which negative triangles do we miss? We can ignore triangles consisting of only unpopular edges as they will be dealt with in future iterations, so only focus on triangles involving at least some popular edge. If such a triangle is just \emph{barely} negative in the sense that
\begin{equation*}
    -\Delta < w(i, k) + w(k, j) + w(i, j) < 0,
\end{equation*}
then the rounding will actually turn this triangle into a zero-weight triangle, and therefore our real-to-integer reduction will correctly take care of it. This only leaves the \emph{very} negative triangles of weight less than $-\Delta$. To deal with these, we follow the ``Fredman's trick meets dominance product'' technique due to Chan, Vassilevska W., and Xu~\cite{ChanVX23}, developed as a follow-up to~\cite{ChanVX22}. The idea is roughly as follows. Each popular edge, say $(i, j)$, is part of many zero-weight triangles. Thus, when sampling a vertex $k'$, with good probability we have that
\begin{equation*}
    w(i, j) = -w(i, k') - w(k', j).
\end{equation*}
Then, the condition that $(i, k, j)$ is a negative triangle can be expressed as
\begin{equation*}
    0 > w(i, k) + w(k, j) + w(i, j) = w(i, k) + w(k, j) - w(i, k') - w(k', j),
\end{equation*}
or, equivalently,
\begin{equation*}
    w(i, k') - w(i, k) > w(k, j) - w(k', j).
\end{equation*}
This elementary rearrangement is sometimes called ``Fredman's trick''~\cite{Fredman}. Note that the left-hand side only depends on $i$ and $k$ and the right-hand side only depends on $k$ and $j$ (assuming that $k'$ is fixed). The problem of detecting a negative triangle therefore reduces to deciding if there is a triangle $(i, k, j)$ in a tripartite graph, where the edge weight in one prescribed edge part is larger than that in another edge part. This is an instance of the so-called \emph{dominance product} problem, which is known to be computable in subcubic time $O(n^{(\omega+3)/2})$~\cite{Matousek91}.\footnote{Earlier we claimed that our reduction is combinatorial and avoids algebraic techniques, but Matou\v{s}ek's algorithm for dominance product relies on fast matrix multiplication. In the actual reduction we instead use the hypothetically fast integer APSP algorithm in place of Matou\v{s}ek's algorithm; this makes the reduction slightly more lossy but combinatorial.} In this way, detecting the very negative triangle only leads to an overall subcubic overhead.

The full real-to-integer reduction ended up being quite technical. See the pseudocode in Algorithm~\ref{alg:apsp-overview} for a high-level summary.

\begingroup
\RestyleAlgo{ruled}
\begin{algorithm}[!t]
    \DontPrintSemicolon
    \SetArgSty{textnormal}
    \SetAlgoNoEnd
    \LinesNumbered
    \caption{Real-to-integer reduction for Negative Triangle}
    \label{alg:apsp-overview}
    \KwIn{A tripartite graph $G = (V,E,w)$ with real edge weights}
    \KwOut{Is there a negative triangle in $G$?}
    \While{$G$ has more than $n^{3-\epsilon}$ triangles}{
        Choose $\Delta$ such that there are $\Theta(n^{3-\epsilon})$ triangles of weight $= 0 \pm \Delta$\;
        \makebox{Partition into \emph{popular} and \emph{unpopular} edges, call a triangle \emph{popular} if it has a popular edge\hspace{-1cm}}\;
        \makebox{Find popular triangles of weight $< -\Delta$ by ``Fredman's trick meets dominance product''}; return YES if there is any\;
        Apply the doubling reduction, preserving popular triangles of weight $= 0\pm\Delta$\;
        Apply the real-to-integer framework to replace popular edges by integers (Freiman's theorem, subset sum covers, Frank--Tardos theorem)\;
        List all $O(n^{3-\epsilon})$ integer non-positive triangles; return YES if any has negative original weight\;
        Remove all popular edges from $E$\;
    }
    List the remaining triangles; return YES if any has negative weight\;
    \Return{NO}\;
\end{algorithm}
\endgroup

\section{Preliminaries} \label{sec:prelims}

\subsection{Notation}
For integer $n \geq 0$, we write $[n] = \set{1, \dots, n}$, and $[\pm n] = \set{-n, \dots, n}$. We write $x = y \pm \Delta$ if $|x - y| \leq \Delta$. We let $X \sqcup Y$ denote the disjoint union of two sets $X$ and $Y$ (i.e., the union when both sets are disjoint). We similarly define disjoint unions of matrices as follows. Suppose that $A, B \in (\Real \cup \set{\bot})^{n \times m}$ are matrices where at each position $A$ or $B$ is equal to the dummy symbol $\bot$. Then we denote by $A \sqcup B$ the matrix obtained from $A$ by replacing the $\bot$-entries by the respective entries in $B$.

We say that an event happens \emph{with high probability} if it happens with probability at least $1 - n^{-c}$, where~$n$ is the input size and where $c$ is a constant which we can choose as large as required.

\subsection{Problems}

\begin{problem}[Integer APSP]
    Given a directed graph with integer edge weights and no negative cycles, compute the length of a shortest path from $u$ to $v$ for every pair of vertices $(u,v)$.
\end{problem}
\begin{problem}[Real APSP]
    Given a directed graph with real edge weights and no negative cycles, compute the length of a shortest path from $u$ to $v$ for every pair of vertices $(u,v)$.
\end{problem}

Using a standard chain of reductions~\cite{VassilevskaW18}, it is well-known that the existence of an algorithm in truly subcubic time for Integer APSP is equivalent to the existence of an algorithm in truly subcubic time for Integer Negative Triangle.
In the same paper, Vassilevska Williams and Williams give another reduction from APSP to Negative Triangle that avoids a binary search on the space of possible triangle weights and hence also works for real numbers.
Therefore, also Real APSP and Real Negative Triangle are subcubically equivalent. In all our reductions we exclusively work with Negative Triangle instead of APSP.

\begin{problem}[Integer Negative Triangle]
    Given three $n \times n$ matrices $A, B, C \in \Int^{n\times n}$, decide if there exist $i,k,j \in [n]$ such that $A[i,k] + B[k,j] + C[i,j] < 0$.
\end{problem}
\begin{problem}[Real Negative Triangle]
    Given three $n \times n$ matrices $A, B, C \in \Real^{n\times n}$, decide if there exist $i,k,j \in [n]$ such that $A[i,k] + B[k,j] + C[i,j] < 0$.
\end{problem}

Exact Triangle is defined the same way with equalities in place of inequalities.

\begin{problem}[Integer Exact Triangle]
    Given three $n \times n$ matrices $A, B, C \in \Int^{n\times n}$, decide if there exist $i,k,j \in [n]$ such that $A[i,k] + B[k,j] + C[i,j] = 0$.
\end{problem}
\begin{problem}[Real Exact Triangle]
    Given three $n \times n$ matrices $A, B, C \in \Real^{n\times n}$, decide if there exist $i,k,j \in [n]$ such that $A[i,k] + B[k,j] + C[i,j] = 0$.
\end{problem}

For the integer variants we throughout assume that the weights are from the universe $\set{-u, \dots, u}$.
Throughout the paper we also allow integer and real instances of any of the defined problem to contain \emph{missing values} replaced by a dummy symbol $\bot$. To simplify notation, we use suitable conventions such as $x + \bot = \bot$ for $x \in \Real$ or $\bot \not < 0$, wherever necessary. Entries equal to $\bot$ could also be simulated by sufficiently large numbers.
\section{Real-to-Integer Toolbox}
\label{sec:toolbox}

In this section, we develop and recap some necessary tools for real-to-integer reductions as well as APSP in general.
We start with some preliminary definitions concerning rounded reals and select-plus rank (\Cref{sec:approx-reals}).
Next, we state the edge weight reduction for APSP that we use, in the spirit of a universe reduction (\Cref{sec:potential-adj}).
Then, we recap our real-to-integer reduction tools and prove the precise primitive used for Exact Triangle and APSP (\Cref{sec:real-to-int-low-rank}).
Finally, we state listing-to-decision reductions (\Cref{sec:list-to-decision}).

\subsection{Approximate Reals and Approximate Select-Plus Rank}\label{sec:approx-reals}

\paragraph{Approximate Reals}
Morally, our algorithm works by rounding real numbers to an appropriate scale to create structure and then uses that structure.
However, the restricted Real RAM model does not allow for a rounding operation, since that would already allow for unreasonably efficient algorithms (e.g.,~\cite{HartmanisS74,Schonhage79}).
Nonetheless, we can carry out the rounding \emph{symbolically}.
The operation we would like to be able to perform is rounding a real $x \in \Real$ to its nearest integer multiple of some scale $\Delta \in \Real$. Instead, we can keep track of $x$ and $\Delta$ separately, and whenever we perform a comparison $\lceil x / \Delta \rfloor = \lceil y / \Delta \rfloor$ of two reals $x, y \in \Real$ that should be rounded, we instead check that $x \in [y-\Delta, y + \Delta]$ (via two comparisons $x \ge y-\Delta$ and $x \le y + \Delta$). 
Note that this is not exactly the same operation as comparing rounded numbers, but performs a similar role in the sense that it allows us to compare numbers on a coarser scale.

We write $x = y \pm \Delta$ to denote $x \in [y-\Delta, y+\Delta]$. 
For two sets $X, C \subseteq \Real$, we say that $C$ is a \emph{$\Delta$-cover} of $X$ if $X \subseteq \bigcup_{c \in C}[c-\Delta, c+\Delta]$. In other words, for every $x \in X$ there should be a $c \in C$ with $x = c \pm \Delta$.
For $\Delta\ge0$, a set $X \subseteq \Real$ is \emph{$\Delta$-spaced} if $|x - x'| \ge \Delta$ for every pair of distinct elements $x,x'\in X$.

We denote by $\cabs{X}$ the size of a minimum $\Delta$-cover of $X$.
Note that such a minimum $\Delta$-cover of size $\cabs{X}$ is easy to compute in time $O(\abs{X} \log \abs{X})$ by a simple greedy algorithm: Indeed, just sort $X$ and sweep over its elements in increasing order. For every encountered and currently uncovered number $x \in X$, we add $x + \Delta$ to the $\Delta$-cover and skip all numbers less than or equal to $x + 2\Delta$. 

\begin{observation}[Computing Minimum $\Delta$-Covers]\label{obs:delta-cover}
    Given $\Delta \in \Realnneg$ and a set $X \subseteq \Real$, we can compute in time $O(\abs X \log |X|)$ a $\Delta$-cover $C$ of $X$ of size $\cabs X$. Additionally, the set $C$ is $2\Delta$-spaced and hence for every $x \in X$ there is at most one element $c \in C$ with $x = c \pm \Delta$.
\end{observation}

\paragraph{Select-Plus Rank for Approximate Reals}
For $\Delta \in \Realnneg$, the \emph{$\Delta$-approximate select-plus rank} (or just \emph{$\Delta$-approximate rank}) of a matrix $A \in (\Real \cup \set{\bot})^{n\times n}$ is the smallest integer $r \ge 0$ such that there are matrices $U \in \Real^{n\times r}$ and $V \in \Real^{r \times n}$ satisfying the following:
For all $(i,j) \in [n]^2$ with $A[i,j] \ne \bot$ there is a $k \in [r]$ such that:
\begin{equation*}
    A[i,j] = U[i,k] + V[k, j] \pm \Delta.
\end{equation*}
When $S \in ([r] \cup \set{\bot})^{n\times n}$ is a matrix such that $S[i,j] = k$, we call $(U,V,S)$ a \emph{$\Delta$-approximate rank-$r$ decomposition} of $A$.
Whenever we say that a matrix $A$ has $\Delta$-approximate rank at most $r$, we assume that we also have access to a $\Delta$-approximate rank-$r$ decomposition of $A$. 

We first observe that the $\Delta$-approximate rank of a matrix is not very sensitive to the choice of $\Delta$, i.e., when we scale $\Delta$ by a constant factor, the rank also only changes by a constant factor.

\begin{observation}\label{obs:change-delta}
    If a matrix $A \in (\Real \cup \set{\bot})^{n \times n}$ has $\Delta$-approximate rank at most $r$, then it also has $\Delta/c$-approximate rank at most $(2c+1)r$ for any $c \in \Intpos$. We can compute it in time $O(n^2c)$.
\end{observation}
\begin{proof}
    Let $(U,V,S)$ be the $\Delta$-approximate rank-$r$ decomposition of $A$.
    To construct a $\Delta/c$-approximate rank-$(2c+1)r$ decomposition, simply replace each $U[i,k]$ with $2c+1$ entries $U[i,k]-c\Delta/(2c), \dots, U[i,k] + c\Delta/(2c)$. Do the same for $V$ and replace $V[k,j]$ with entries  $V[k,j]-c\Delta/(2c), \dots, V[k,j] + c\Delta/(2c)$.
    Finally, update $S$ by finding the correct new entry among the $2c+1$ newly created entries.
    This also dominates the running time.
\end{proof}

\subsection{Potential-Adjusting Reductions and Edge Weight Reduction}\label{sec:potential-adj}

To make it easier to keep track of how our reductions manipulate the graph and the number of triangles of certain weights, we reuse the definition of a \emph{potential-adjusting reduction}, defined in~\cite{Fischer26}.
Informally, a potential-adjusting reduction is a reduction that allows us to delete edges and to add node weights in a way such that all triangles of weight 0 (up to $\pm \Delta$ in our case) are kept track of.
Moreover, every triangle that survives the reduction will keep its weight (the value of $A[i,k] + B[k,j] + C[i,j]$) unchanged.
We use the following two definitions.

\begin{definition}[Potential Adjustment]
    Let $(A,B,C), (A', B', C')$ be instances of Real Exact Triangle of the same size $n$.
    We say that $(A',B',C')$ is a \emph{potential adjustment} of $(A,B,C)$ if there are \emph{potential functions} $u,v,w \in \Real^n$ such that:
    \begin{equation*}
        \begin{array}{r@{~}r@{\quad}r@{\quad}r@{~}r@{\quad}l@{~}l}
            A'[i,k] =& A[i,k]-u[i]+v[k] &\text{or}& A'[i,k]=&\perp &\text{for all }(i,k)&\in[n]^2,\\[1.5mm]
            B'[k,j] =& B[k,j]-v[k]+w[j] &\text{or}& B'[k,j]=&\perp &\text{for all }(k,j)&\in[n]^2,\\[1.5mm]
            C'[i,j] =& C[i,j]+u[i]-w[j] &\text{or}& C'[i,j]=&\perp &\text{for all }(i,j)&\in[n]^2.
        \end{array}
    \end{equation*}
\end{definition}

\begin{definition}[$\Delta$-Approximate Potential-Adjusting Reduction]
A \emph{$\Delta$-approximate potential-adjusting reduction} in time $T$ is an algorithm that runs in time $T$ and takes as input a parameter $\Delta \in \Realnneg$ as well as a Real Exact Triangle instance $(A,B,C)$ of size $n$ and outputs a pair $(\mathcal I, \mathcal T)$ where
\begin{itemize}
    \item $\mathcal I$  is a set of potential adjustments of $(A,B,C)$, and
    \item $\mathcal T \subseteq [n]^3$ is a set of triples of size $\abs{\mathcal T} \le T$,
\end{itemize}
such that each $\Delta$-approximate triangle $(i,k,j)$ in $(A,B,C)$ appears in $\mathcal T$ or as a $\Delta$-approximate triangle in an instance in $\mathcal I$.
\end{definition}

We use the following edge weight reduction to an approximately uniform and approximately low doubling weight set for matrices of bounded approximate rank.
The statement (and reduction chain in the proof) mostly mirrors its exact version in \cite{Fischer26} (some of which already mirror reductions from~\cite{AbboudFJVX25} in a rectangular setting).
Our approach inherently only works for the corresponding $\Delta$-approximate definitions, which means we have to reprove the weight reduction for this setting. 
This is mostly a tedious but natural transfer to corresponding approximate versions of each argument in the chain.
The proof of \Cref{lem:low-rank-to-small-doubling} can be found in \Cref{sec:proof-low-rank-to-small-doubling}.
It also includes a port of the algorithmic version of the famous BSG theorem~\cite{BalogS94,Gowers01,ChanL15}, to a version for approximate reals, see \Cref{thm:bsg_approx}.

Recall that for $\Delta \in \Realnneg$ and a set $X \subseteq \Real$, we denote by $\cabs X$ the size of the minimum $\Delta$-cover, that is, the minimum number of width-$2\Delta$ intervals necessary to cover all of $X$.

\begin{restatable}[Reduction to Approximately Uniform and Low-Doubling Weights]{lemma}{weightreduction}\label{lem:low-rank-to-small-doubling}
    Let $s \ge 1$, $\Delta \in \Realnneg$ and let $(A,B,C)$ be an Exact Triangle instance where one of the matrices has $\Delta$-approximate rank at most $r$.
    
    There is a $\Delta$-approximate potential-adjusting reduction to $sn^{o(1)}$ instances where for each instance the set of weights $X \subseteq \Real$ satisfies:
    \begin{itemize}
        \item \emph{(Approximate Uniformity)} $\cabs X \le r$, and
        \item \emph{(Approximate Low-Doubling)} $\cabs{X+X} \le s\cdot \cabs X \le sr$.
    \end{itemize}
    The reduction runs in time \[\parens*{rsn^{2} + \frac{n^3 }{ s^{1/10^5}}}^{1+o(1)}.\]
\end{restatable}

\subsection{Real-to-Integer Reduction for Low-Rank Reals}\label{sec:real-to-int-low-rank}

In this subsection, we first list the necessary real-to-integer toolbox, which we reuse from~\cite{real3sum}.

\begin{definition}[Subset Sum Cover] \label{def:subset-sum-cover}
    Let $A \subseteq \Real$. A pair $(\alpha, f)$, where $\alpha \in \Real^d$ and $f : A \to \set{0, 1}^d$, is called a \emph{$\Delta$-approximate subset sum cover} of $A$ if it satisfies $a = \alpha^T f(a) \pm \Delta$ for each $a \in A$. We call $d$ the \emph{rank} of the subset sum cover.
\end{definition}

\begin{theorem}[Poor Man's Freiman's Theorem~\protect{\cite[Theorem 3.10]{real3sum}}]\label{thm:poor-freiman}
    Let $\Delta \in \Realnneg$, $K \ge 1$, and let $A \subseteq \Real$ be a set satisfying the $\Delta$-approximate doubling condition $\cabs{A+A} \le K\cabs A$.
    Then there is a $\Delta$-approximate subset sum cover of $A$ with rank $(K \log|A|)^{O(1)}$. 
    It can be computed in time $\tilde O(|A|^2)$.
\end{theorem}

The next lemma is also already proven in~\cite{real3sum}, except for the third property. Since the hash function in the proof is monotone except for a random offset independent of the function argument, property (iii) follows immediately from property (ii).
The proof internally uses the Frank--Tardos theorem~\cite{FrankT87}.

\begin{lemma}[Hashing Reals to Integers~\protect{\cite[Lemma~3.9]{real3sum}}]\label{lem:hash-popular-to-int}
    Let $\Delta \in \Realnneg$ and let~$A \subseteq \Real$ be a set with a $\Delta$-approximate subset sum cover of rank $(K\log |A|)^{O(1)}$.
    There is a randomized algorithm that runs in time $\tilde O(|A| + (K\log|A|)^{O(1)})$ and computes a function $h : A \to [\pm 2^{(K\log |A|)^{O(1)}}]$ such that:
    \begin{enumerate}[label=(\roman*)]
        \item For all $a, b, c \in A$ with $a + b + c = 0 \pm \Delta$, we have $\Pr[h(a) + h(b) + h(c) = 0] \geq \Omega(1)$.
        \item For all $a, b, c \in A$ with $h(a) + h(b) + h(c) = 0$, we have $a + b + c = 0 \pm 100 \Delta$.
        \item For all $a,b,c \in A$ with $h(a) + h(b) + h(c) < 0$, we have $a+b+c < 100\Delta$. 
    \end{enumerate}
\end{lemma}

The next lemma is the main goal of this subsection.
It combines the previously cited tools and is the single way in which we use them in the main proof.
The lemma shows that if at least one input matrix has low $\Delta$-approximate rank, we can convert the edge weights to integers.

\begin{lemma}[Real-to-Integer Reduction for Low-Rank Reals]\label{lem:unified-popular-case}
    Let $\Delta \in \Realnneg$, let $s \ge 1$ be a parameter, and let $A,B,C \in (\Real \cup \set{\bot})^{n\times n}$ such that:
    \begin{enumerate}
        \item at least one of $A$, $B$, and $C$ has $\Delta$-approximate rank at most $r$, and
        \item the number of $2\Delta$-approximate exact triangles in $(A,B,C)$ is at most $T$.
    \end{enumerate}
    Then there is an algorithm that reports a set of triangles $\mathcal T$, and outputs $sn^{o(1)}$ triples of integer matrices $X,Y,Z \in ([\pm 2^{(s \log r)^{O(1)}}] \cup \set{\bot})^{n \times n}$ such that:
    \begin{enumerate}[label=(\roman*)]
        \item $|\mathcal T| \leq T$, and the number of exact triangles in any $(X,Y,Z)$ is at most $T$, 
        \item any $\Delta$-approximate exact triangle in $(A,B,C)$ is in $\mathcal T$ or an exact triangle in some~$(X,Y,Z)$,
        \item any negative triangle $(i,k,j)$ in some instance $(X,Y,Z)$ satisfies $A[i,k] + B[k,j] + C[i,j] < 2\Delta$.
    \end{enumerate}
    The reduction runs in time \[\parens*{ (rs)^{O(1)}n^2 + \frac{n^3}{s^{1/10^5}}}^{1+o(1)}.\]
\end{lemma}

\begin{proof}
    Assume that we are given a $\Delta$-approximate rank-$r$ decomposition of $C$, otherwise exchange the matrices appropriately.
    First, we use \Cref{obs:change-delta} to turn it into a $\Delta'$-approximate rank decomposition of rank $r' \coloneqq 1001r$ for $\Delta' \coloneqq \Delta/500$. Initialize $\mathcal T\coloneqq\emptyset$. For each shift $\sigma \in \Delta' \cdot \set{-500, -499, \ldots, 500}$, run the following steps on $(A, B, C_\sigma)$, where $C_\sigma[i, j] = C[i, j] - \sigma$ for non-$\bot$ entries. Subtracting $\sigma$ appropriately in the low-rank decomposition gives a $\Delta'$-approximate rank-$r'$ decomposition of $C_\sigma$.
    Then apply the edge weight reduction to approximately uniform and small doubling weights (\Cref{lem:low-rank-to-small-doubling}) at scale $\Delta'$ to $(A,B,C_\sigma)$. This returns a pair $(\mathcal I, \mathcal T_\sigma)$ of $\abs{\mathcal I} \le sn^{o(1)}$ potential-adjusted instances and some exceptional triangles $\mathcal T_\sigma$. Insert into $\mathcal T$ every triangle $(i,k,j) \in \mathcal T_\sigma$ whose original weight satisfies $A[i, k] + B[k, j] + C[i, j] = 0 \pm \Delta$.
    
    Next, we process the instances $(\hat A,\hat B,\hat C) \in \mathcal I$. Let $X \subseteq \Real$ denote the set of entries in the three matrices~$\hat A, \hat B, \hat C$. This set satisfies $ |X|_{\pm \Delta'} \le r'$ and $|X+X|_{\pm\Delta'} \le s |X|_{\pm\Delta'}$.
    Compute the greedy minimum $\Delta'$-cover $Y$ of~$X$, of size $|Y|=|X|_{\pm\Delta'}\le r'$ (with \Cref{obs:delta-cover}) and for $x \in X$ write $f(x)$ for the unique $y \in Y$ with $x = y \pm \Delta'$.
    The greedy construction gives $Y\subseteq X+\Delta'$ and places its centers more than $2\Delta'$ apart. Thus~$|Y|_{\pm\Delta'}=|Y|$, and therefore
    \[
        |Y+Y|_{\pm\Delta'}\le |X+X|_{\pm\Delta'}\le s|X|_{\pm\Delta'}=s|Y|_{\pm\Delta'}.
    \]
    Apply Poor Man's Freiman's theorem to $Y$ to compute a $\Delta'$-approximate subset sum cover of rank $(s\log r')^{O(1)}$, which is also a $4\Delta'$-approximate subset sum cover.

    For $g = 1, \dots, O(\log n)$, perform the following steps: Hash the reals in $Y$ to integers with \Cref{lem:hash-popular-to-int} at scale $4\Delta'$, and let \smash{$h_g: Y \to [\pm 2^{(s\log r)^{O(1)}}]$} denote the resulting hash function. 
    Replace all occurrences in~$\hat A$,~$\hat B$, and~$\hat C$ of a real $x\in \Real$ with the integer $h_g(f(x))$, and add the resulting integer instance to the output. In the end, return $\mathcal T$ together with the integer instances.

    \paragraph{Correctness.} 
    Consider a $\Delta$-approximate exact triangle in the original instance with weight $w := A[i, k] + B[k, j] + C[i, j]$. Focus on the shift $\sigma$ satisfying $|w - \sigma| \le \Delta'$. For this shift, the triangle is either in $\mathcal T_\sigma$, in which case it is added to $\mathcal T$, or it satisfies 
    \begin{equation*}
        \hat A[i,k] + \hat B[k,j] + \hat C[i,j] = 0 \pm\Delta'
    \end{equation*}
    for at least one instance $(\hat A, \hat B, \hat C) \in \mathcal I$ by the guarantee of the $\Delta'$-approximate potential-adjusting reduction.
    Then
    \begin{equation*}
        f(\hat A[i,k]) + f(\hat B[k,j]) + f(\hat C[i,j]) = 0 \pm 4\Delta',   
    \end{equation*}
    since $f(x) = x \pm \Delta'$ for all $x \in X$. For any $g$, it also satisfies 
    \begin{equation*}
        h_g(f(\hat A[i,k])) + h_g(f(\hat B[k,j])) + h_g(f(\hat C[i,j])) = 0
    \end{equation*}
    with constant probability by \Cref{lem:hash-popular-to-int}, property (i), making $(i,k,j)$ an exact integer triangle in the reduced instance.
    After $O(\log n)$ independent repetitions, a union bound over the at most $n^3$ triangles shows that every $\Delta$-approximate exact triangle outside $\mathcal T$ appears as an exact triangle in at least one integer instance with high probability, proving property (ii).

    We also need to argue that we do not get too many integer exact triangles, that is, property (i). First, every triangle in $\mathcal T$ is $\Delta$-approximate in the original instance, so $|\mathcal T|\le T$.
    Next, use property (ii) of the real-to-integer hash function (\Cref{lem:hash-popular-to-int}), which states that any integer triangle with 
    \begin{equation*}
        h_g(f(\hat A[i,k])) + h_g(f(\hat B[k,j])) + h_g(f(\hat C[i,j])) = 0,
    \end{equation*}
    for some $g$, also satisfies
    \begin{equation*}
        f(\hat A[i,k]) + f(\hat B[k,j]) + f(\hat C[i,j]) = 0 \pm 400\Delta'.
    \end{equation*}
    Since $f(x) = x \pm \Delta'$ for all $x \in X$, we must also have
    \begin{equation*}
        \hat A[i,k] + \hat B[k,j] + \hat C[i,j] = 0 \pm 500\Delta'.
    \end{equation*}
    Potential adjustments preserve this triangle weight; undoing the shift only adds $\sigma$. Thus the original weight has magnitude at most $|\sigma| + 500\Delta' \le 2\Delta$. There are at most $T$ such triangles.

    We prove the third property about negative triangles in a similar way. Consider $(i,k,j)$ and some $g$ such that 
    \begin{equation*}
        h_g(f(\hat A[i,k])) + h_g(f(\hat B[k,j])) + h_g(f(\hat C[i,j])) < 0.
    \end{equation*}
    Then
    \begin{equation*}
        \hat A[i,k] + \hat B[k,j] + \hat C[i,j] < 500\Delta',
    \end{equation*}
    so the original weight is less than $\sigma+500\Delta'\le2\Delta$, proving property (iii).

    \paragraph{Running Time.}
    The edge weight reduction (\Cref{lem:low-rank-to-small-doubling}) runs in time $(rsn^2 + n^3/s^{1/10^5})^{1+o(1)}$ per shift, which also bounds the number of exceptional triangles that we enumerate. Computing $Y$ and the map $f$ takes time $\tilde O(n^2)$ per instance. Poor Man's Freiman's theorem (\Cref{thm:poor-freiman}) takes time $\tilde O(r^2)$ per instance.
    Finally, hashing each instance to integers (\Cref{lem:hash-popular-to-int}) takes time $\tilde O(r + (s\log r)^{O(1)})$ per instance.
    There are $O(1)$ shifts and we repeat $O(\log n)$ times. In total, the running time becomes
    \begin{equation*}
        \parens*{ (rs)^{O(1)}n^2 + \frac{n^3}{s^{1/10^5}} }^{1+o(1)},
    \end{equation*}
    as claimed.
\end{proof}

\subsection{Listing-to-Decision Reductions for Triangle Problems}\label{sec:list-to-decision}

We use the following standard listing-to-decision reductions for the triangle problems of interest, Exact Triangle and Negative Triangle, due to Vassilevska Williams and Williams~\cite{VassilevskaW18}. 
They show that an algorithm that solves the problem in subcubic time can also be used to list a subcubic number of solutions in subcubic time.

\begin{lemma}[Listing-to-Decision for Integer Exact Triangle, \protect{\cite{VassilevskaW18}}]\label{lem:exact-triangle-listing}
    If Integer Exact Triangle can be decided in time $O(n^{3-\epsilon} (\log u)^{O(1)})$ for some constant $\epsilon > 0$, then one can also list up to $T$ solutions in time $O(n^{3 - \epsilon} T^{\epsilon/3} (\log u)^{O(1)})$.
\end{lemma}

\begin{lemma}[Listing-to-Decision for Integer Negative Triangle, \protect{\cite{VassilevskaW18}}]\label{lem:negative-triangle-listing}
    If Integer Negative Triangle can be decided in time $O(n^{3-\epsilon} (\log u)^{O(1)})$ for some constant $\epsilon > 0$, then one can also list up to $T$ solutions in time $O(n^{3 - \epsilon} T^{\epsilon/3} (\log u)^{O(1)})$.
\end{lemma}

\section{Real-to-Integer Reduction for Exact Triangle} \label{sec:exact-triangle}

In this section we prove \Cref{thm:exact-triangle} and develop almost all remaining parts to prove our real-to-integer reduction for APSP (\Cref{thm:apsp}).

\subsection{Popular/Unpopular Decomposition}

The initial setup of the algorithm is more or less exactly the same as for 3SUM~\cite{real3sum}; the adaptations to triangles instead of 3SUM solutions are mainly concerned with the different number of possible solutions.

We call a triple $(i, k, j) \in [n]^3$ a \emph{$\Delta$-approximate triangle} (or \emph{$\Delta$-approximate solution}) if $A[i, k] + B[k, j] + C[i,j] = 0 \pm \Delta$.

\begin{lemma}[Approximation Threshold]\label{lem:threshold}
    Let $A,B,C \in (\Real \cup \set{\bot})^{n\times n}$ be matrices and let $K \ge 1$.
    There is a randomized algorithm that runs in time $K(\log n)^{O(1)}$ and either detects an Exact Triangle solution (in which case there are at least $\Omega(n^3/K)$ exact solutions), decides that the number of $\infty$-approximate solutions is at most $O(n^3/K)$, or computes a threshold $\Delta \in \Real_{\ge 0}$ such that:
    \begin{enumerate}[label=(\roman*)]
        \item the number of $\Delta$-approximate solutions is at most $O(n^3 / K)$, and
        \item the number of $2\Delta$-approximate solutions is at least $\Omega(n^3 / K)$.
    \end{enumerate}
\end{lemma}
\begin{proof}
    Consider a sequence of length $n^3$ of numbers equal to the absolute weight of all triangles $|A[i,k] + B[k,j] + C[i,j]|$ for all triples $i,k,j \in [n]$, sorted in increasing order (where $\bot$-values are considered the largest values).
    We aim to find one triple $(i^*, k^*, j^*)$ whose rank in this sorted order is between $n^3/K$ and $2n^3/K$.
    After having found such a triple we can return one of the three possible answers:
    If $A[i^*,k^*] + B[k^*, j^*] + C[i^*,j^*] = 0$, we have found an Exact Triangle solution. In this case, the rank of the selected triple also guarantees at least $n^3/K$ exact solutions.
    If one of $A[i^*,k^*]$, $B[k^*, j^*]$, and $C[i^*,j^*]$ equals $\bot$, then so does their sum and we can conclude that the number of $\infty$-approximate solutions is at most $2n^3/K$.
    If none of these cases apply, we set $2\Delta \coloneqq |A[i^*,k^*] + B[k^*, j^*] + C[i^*,j^*]|$. Since the rank of $(i^*,k^*,j^*)$ is at least $n^3/K$, there must be at least $n^3/K$ triples $(i,k,j)$ with
    \begin{equation*}
        |A[i,k] + B[k,j] + C[i,j]| \le |A[i^*,k^*] + B[k^*, j^*] + C[i^*,j^*]| = 2\Delta.
    \end{equation*}
    All of these triples form $2\Delta$-approximate solutions. On the other hand, there are at most $2n^3/K$ triples $(i,k,j)$ with 
    \begin{equation*}
        |A[i,k] + B[k,j] + C[i,j]| < |A[i^*,k^*] + B[k^*, j^*] + C[i^*,j^*]| = 2\Delta,
    \end{equation*}
    and only those triples can form $\Delta$-approximate solutions since $\Delta > 0$.

    To find $(i^*, k^*, j^*)$ we apply a sketching technique. Let $R \coloneqq \Theta(\log n)$ and sample $R K$ triples $(i, k, j) \in [n]^3$ uniformly at random. Among them, pick the one for which $|A[i,k] + B[k,j] + C[i,j]|$ is $1.5R$-th smallest. Equivalently, we choose $R K$ uniformly random positions in the sorted order, and commit to the $1.5R$-th smallest position among the samples. For each sample, the probability that it has rank below $n^3 / K$ is exactly $1/K$. Therefore, the expected number of samples below this rank is $R K / K = R$. By Chernoff's bound it follows that with high probability there are fewer than $1.5R$ samples below rank $n^3 / K$ and the lower bound on the $2\Delta$-approximate solutions follows. Similarly, the expected number of samples below rank $2n^3 / K$ is $2R$, and by Chernoff's bound with high probability there are at least $1.5R$ such samples. Thus, the upper bound on the $\Delta$-approximate solutions holds too.
\end{proof}

\begin{lemma}[Popular/Unpopular Decomposition]\label{lem:decomp-pop}
    Let $A,B,C \in (\Real \cup \set{\bot})^{n\times n}$ be matrices, $\Delta \in \Realnneg$, and let $L \ge 1$.
    There is a randomized algorithm that runs in time $\tilde O(n^2L^2)$ and partitions the edge set into \emph{popular} and \emph{unpopular} edges such that:
    \begin{enumerate}[label=(\roman*)]
        \item Each popular edge is in at least $n/L$ $2\Delta$-approximate solutions.
        \item There are at most $6n^3/L$ $2\Delta$-approximate solutions involving unpopular edges.
    \end{enumerate}
\end{lemma}
\begin{proof}
    We first describe a simple deterministic but slow solution:
    Initially, treat every edge as popular. 
    For each edge that is contained in fewer than $n/L$ $2\Delta$-approximate solutions, we mark it as unpopular.
    After this process ends, the first property holds by construction.
    By definition, each unpopular edge is involved in at most $n/L$ $2\Delta$-approximate solutions. Summing over all $3n^2$ edges proves property (ii).

    Since we cannot afford to compute the number of triangles exactly, we implement the above idea with estimated counts.
    For each $C$-edge $(i,j) \in [n]^2$, sample $\Theta(L^2 \log n)$ uniformly random values $k \in [n]$ and count how many such triples form a $2\Delta$-approximate solution $(i,k,j)$. ($A$ and $B$ are symmetric.)
    We can then simply scale the sample count to get an additive approximation of the number of solutions involving this edge within error $n/2L$. Chernoff's bound shows that the approximation works with high probability. 

    Now simply use the estimated counts to run the above described algorithm but mark an edge as unpopular if its estimated number of $2\Delta$-approximate solutions is less than the threshold $\tau \coloneqq 3n/2L$.
    After the algorithm, each popular edge is still in at least $\tau -n/2L = n/L$ $2\Delta$-approximate solutions.
    Summing over all edges shows that the number of $2\Delta$-approximate solutions involving unpopular edges is at most 
    \begin{equation*}
        3n^2 \cdot \parens*{\tau + \frac{n}{2L}} = 3n^2 \cdot \frac{2n}{L} = \frac{6n^3}{L}. \qedhere
    \end{equation*}
\end{proof}

Now, we are left with only 3 types of solutions:
\begin{enumerate}
    \item Solutions with only popular edges,
    \item solutions with at least one popular edge and at least one unpopular edge, and
    \item solutions with only unpopular edges.
\end{enumerate}
We show how to detect any solutions of types 1 and 2 in \Cref{sec:only-pop,sec:pop-unpop} respectively. In \Cref{sec:only-unpop}, we show how to deal with solutions of type 3 and complete the proof of \Cref{thm:exact-triangle}.

Combining the lower bound on the number of $2\Delta$-approximate solutions from the choice of $\Delta$ (\Cref{lem:threshold}) and the upper bound on the number of $2\Delta$-approximate solutions involving unpopular elements, we can observe the following.
\begin{observation}[Number of Approximate Solutions]\label{obs:number-solutions}
    Assume that $L$ is larger than $K$ by a sufficiently large constant factor. Then:
    \begin{enumerate}[label=(\roman*)]
        \item the number of $2\Delta$-approximate solutions of type 1 is at least $\Omega(n^3 / K)$,
        \item the number of $2\Delta$-approximate solutions of types 2 and 3 is at most $O(n^3 / L)$.
    \end{enumerate}
\end{observation}

Recall the definition of the $\Delta$-approximate rank of a matrix.
We prove that after applying our decomposition into popular and unpopular edges, the popular edges must have low rank.

\begin{lemma}[Popular Instances Have Low Rank]\label{lem:popular-low-rank}
    Let $\Delta \in \Realnneg$, $L\ge 1$, and $A,B,C \in (\Real \cup \set{\bot})^{n\times n}$ be matrices where each $C$-edge $(i,j)$ with $C[i,j] \ne \bot$ is in at least $n/L$ $\Delta$-approximate solutions. Then, we can compute a $\Delta$-approximate rank-$O(L\log n)$ decomposition of $C$ in time $O(Ln^2\log n)$.
\end{lemma}
\begin{proof}
    Initialize the $\Delta$-approximate rank-$O(L \log n)$ decomposition $(U,V,S)$ such that $U$ and $V$ contain only $0$ and $S$ contains only $\bot$ entries.
    The algorithm runs in $\Theta(L \log n)$ iterations.
    In an iteration $g$, we sample $k \in [n]$ uniformly at random.
    Now, for any $(i,j) \in [n]^2$ such that $(i,k,j)$ is a $\Delta$-approximate solution, we set 
    \begin{equation*}
        U[i, g] \coloneqq -A[i,k], \quad V[g,j] \coloneqq -B[k,j], \quad \text{and } S[i,j] \coloneqq g.
    \end{equation*}
    Finally, we return $(U,V,S)$ as our decomposition.

    \paragraph{Correctness.}
    Suppose $k$ is sampled in iteration $g$. For all $i,j$ for which $(i,k,j)$ is a $\Delta$-approximate solution, that is, 
    \begin{equation*}
        A[i,k] + B[k,j] + C[i,j] = 0 \pm \Delta,
    \end{equation*}
    we have 
    \begin{equation*}
        U[i,g] + V[g,j] = -A[i,k] - B[k,j] = C[i,j] \pm \Delta.
    \end{equation*}
    Therefore, $S[i,j]$ is assigned correctly.
    
    It remains to show that after $\Theta(L \log n)$ iterations, every $S[i,j]$ was assigned at least once.
    Since every $(i,j)$ is in at least $n/L$ $\Delta$-approximate solutions, every iteration assigns $S[i,j]$ with probability at least $1/L$.
    Hence after $O(L\log n)$ iterations, every $(i,j)$ was assigned at least once with high probability and we return a correct $\Delta$-approximate rank-$O(L\log n)$ decomposition of $C$.

    \paragraph{Running Time.}
    Each iteration can easily be implemented in time $O(n^2)$. The running time therefore follows from the number of iterations.
\end{proof}

\subsection{Type 1: Only Popular Edges}
\label{sec:only-pop}

\begin{lemma}[Type-1 Solutions]\label{lem:type1}
    Assume Integer Exact Triangle is in time $O(n^{3-\epsilon} (\log u)^{O(1)})$. Then there is an algorithm that detects any solution with only popular edges in time \[\parens*{ n^2(KL)^{O(1)} + \frac{|P|n^{1+o(1)}}{K^{\Omega(\epsilon)}} }^{1+o(1)}.\]
\end{lemma}
\begin{proof}
    First set the unpopular entries of $C$ to $\bot$, keeping $A$ and $B$ unchanged. 
    By the definition of $P$ (\Cref{lem:decomp-pop}) and \Cref{lem:popular-low-rank}, we can compute a $2\Delta$-approximate rank-$O(L\log n)$ decomposition for $C$, which we can easily turn into a $\Delta/2$-approximate rank-$O(L\log n)$ decomposition with \Cref{obs:change-delta}. Now also set the unpopular entries of $A$ and $B$ to $\bot$. There are only $|P|$ edges left.

    We apply a standard high degree/low degree split. For some parameter $M$ to be determined later, call a node $v$ \emph{light} if $\deg(v) \le M$ and \emph{heavy} otherwise.
    We enumerate all triangles involving at least one light node by iterating over all light nodes $v$ and all pairs of adjacent edges. The total number of enumerated triangles is 
    \begin{equation*}
        \sum_{v : \deg(v) \le M} \deg(v)^2 \le M\sum_v \deg(v) = 2M|P|.
    \end{equation*}

    The remaining triangles to consider consist of only heavy nodes, of which there are at most $2|P|/M$. Restrict the graph (and the matrices $A$, $B$, and $C$) by removing all light nodes. Hence, we can assume that the matrices $A$, $B$, and $C$ have at most $2|P|/M$ rows and columns. We can also restrict the $\Delta/2$-approximate rank-$O(L\log n)$ decomposition accordingly. 
    
    Also, by the choice of $\Delta$ (\Cref{lem:threshold}), there are at most $T=O(n^3/K)$ $\Delta$-approximate triangles.
    Hence, we can apply the reduction to integer listing (\Cref{lem:unified-popular-case}) at scale $\Delta/2$ with a parameter $s$ to be specified later to reduce to $sn^{o(1)}$ instances of listing up to $T$ solutions to Integer Exact Triangle with integer weights in \smash{$[\pm 2^{(s\log (L\log n))^{O(1)}}]$}.
    We solve all of these instances using the listing-to-decision reduction (\Cref{lem:exact-triangle-listing}). Check the original real weight of every triangle in the returned set $\mathcal T$ and every listed integer triangle, and return an exact solution if one is found.

    \paragraph{Running Time.}
    Computing the rank-decomposition takes time $O(Ln^2\log n)$. 
    Enumerating all triangles involving light nodes takes time $O(M|P|)$.
    Applying the reduction to integer listing (\Cref{lem:unified-popular-case}) takes time $((Ls)^{O(1)}(|P|/M)^2 + (|P|/M)^3/s^{1/10^5})^{1+o(1)}$.
    Finally, listing all integer exact triangles (\Cref{lem:exact-triangle-listing}) takes time $O((|P|/M)^{3-\epsilon}T^{\epsilon/3}(s\log (L \log n))^{O(1)})$ per instance.
    In total, the time is hence
    \begin{equation*}
        \parens*{ M|P| + \frac{(Ls)^{O(1)}|P|^2}{M^2} + \frac{|P|^3}{M^3s^{1/10^5}} + \frac{|P|^{3-\epsilon}n^{\epsilon + o(1)}s^{O(1)}}{M^{3-\epsilon} K^{\epsilon/3}} }^{1+o(1)}.
    \end{equation*}
    Set $M = n / s^{1/10^6}$ and use $|P| \le 3n^2$ and hence $|P|^2/M^3 \le O(ns^{3/10^6})$ and $|P|^{3-\epsilon} \le O(|P|n^{4-2\epsilon})$ to bound the running time by
    \begin{equation*}
        \parens*{ n^2(Ls)^{O(1)} + \frac{|P|n}{s^{\Omega(1)}} + \frac{|P|n^{1+o(1)} s^{O(1)}}{K^{\epsilon/3}} }^{1+o(1)}.
    \end{equation*}
    By choosing $s = K^{\Theta(\epsilon)}$ sufficiently small, the second term dominates the third and we arrive at the final running time 
    \begin{equation*}
        \parens*{ n^2(KL)^{O(1)} + \frac{|P|n^{1+o(1)}}{K^{\Omega(\epsilon)}} }^{1+o(1)}.\qedhere
    \end{equation*}
\end{proof}

\subsection{Type 2: Popular and Unpopular Edges}
\label{sec:pop-unpop}

\begin{lemma}[Type-2 Solutions]\label{lem:type2}
    Assume Integer Exact Triangle is in time $O(n^{3-\epsilon} (\log u)^{O(1)})$. Then there is an algorithm that detects any solution with at least one popular and at least one unpopular edge in time \[\parens*{ L^{O(1)}n^2 + \frac{n^{3+o(1)}}{L^{\Omega(\epsilon)}} }^{1+o(1)}.\]
\end{lemma}
\begin{proof}
    Without loss of generality suppose that we look for triangles that have a popular edge in $C$. By exchanging the roles of $A$, $B$, and $C$, we can cover all cases of where the popular edge lies.
    Hence, we can set any entry in $C$ to $\bot$ if it is not in $P$.
    By the definition of $P$ (\Cref{lem:decomp-pop}) and \Cref{lem:popular-low-rank}, we can compute a $2\Delta$-approximate rank-$O(L\log n)$ decomposition for $C$. Refine it to scale $\Delta/2$ using \Cref{obs:change-delta}, keeping rank $O(L\log n)$. Then run the following steps separately on $(A_U,B,C)$ and $(A,B_U,C)$, where $A_U$ and $B_U$ retain only unpopular entries. Each instance contains an unpopular edge in every triangle and inherits the decomposition of $C$.
    
    There are only $T = O(n^3/L)$ $2\Delta$-approximate triangles with an unpopular edge (\Cref{obs:number-solutions}).
    Hence, we can apply the reduction to integer listing (\Cref{lem:unified-popular-case}) at scale $\Delta/2$ with a parameter $s$ to be specified later to reduce to $sn^{o(1)}$ instances of listing up to $T$ solutions to Integer Exact Triangle with integer weights in $[\pm 2^{(s\log (L \log n))^{O(1)}}]$. We solve all of these instances using the listing-to-decision reduction (\Cref{lem:exact-triangle-listing}). Check the original real weight of every triangle in the returned set $\mathcal T$ and every listed integer triangle, and return an exact solution if one is found.

    \paragraph{Running Time.}
    Computing the rank-decomposition takes time $O(Ln^2\log n)$.
    The reduction to integer listing (\Cref{lem:unified-popular-case}) takes time $((Ls)^{O(1)}n^2 + n^3/s^{1/10^5})^{1+o(1)}$.
    Listing all integer solutions (\Cref{lem:exact-triangle-listing}) takes time $O(n^{3-\epsilon}T^{\epsilon/3}(s\log (L\log n))^{O(1)})$ per instance.
    In total, the running time is
    \begin{equation*}
        \parens*{ (Ls)^{O(1)}n^2 + \frac{n^3}{s^{1/10^5}} + \frac{n^{3+o(1)}s^{O(1)}}{ L^{\epsilon/3}} }^{1+o(1)}.
    \end{equation*}
    In contrast to type-1 solutions (\Cref{lem:type1}), we set $s = L^{\Theta(\epsilon)}$ sufficiently small such that $1/s^{1/10^5}$ dominates $s^{O(1)}/L^{\epsilon/3}$. The final running time then becomes
    \begin{equation*}
        \parens*{ L^{O(1)}n^2 + \frac{n^{3+o(1)}}{L^{\Omega(\epsilon)}} }^{1+o(1)}.\qedhere
    \end{equation*}
\end{proof}

\subsection{Type 3: Only Unpopular Edges}
\label{sec:only-unpop}

\begin{proof}[Proof of \Cref{thm:exact-triangle}]
    We proceed in iterations. In one iteration, we first compute a threshold $\Delta$ (or directly detect a solution and return YES) (\Cref{lem:threshold}). Then, we compute the partition $E = P \sqcup U$ into popular and unpopular edges (\Cref{lem:decomp-pop}).
    Next, we check for solutions involving at least one edge in $P$ (\Cref{lem:type1,lem:type2}). 
    After that, we can remove all edges in $P$ by setting the corresponding matrix entries to $\bot$.
    Finally, when \Cref{lem:threshold} reports that there are at most $O(n^3/K)$ $\infty$-approximate solutions, we list them in time $O(n^3/K^{\epsilon/3})$ with the listing-to-decision reduction (\Cref{lem:exact-triangle-listing}), by first replacing non-$\bot$ entries in the real instance by the integer $0$.

    Denote by $I$ the number of iterations and $P_i$ the set of popular edges in iteration $i$. We can bound the final running time by
    \begin{equation*}
        O\parens*{ \frac{n^3}{K^{\epsilon/3}} } + \sum_{i=1}^I \parens*{ n^2(KL)^{O(1)} + \frac{|P_i|n^{1+o(1)}}{K^{\Omega(\epsilon)}} + \frac{n^{3+o(1)}}{L^{\Omega(\epsilon)}} }^{1+o(1)}.
    \end{equation*}

    It remains to bound $I$. 
    To do so, remember that the number of $2\Delta$-approximate solutions involving only popular edges is at least $\Omega(n^3/K)$ (\Cref{obs:number-solutions}).
    But in any graph with $m$ edges, the number of triangles is at most $O(m^{3/2})$, by a simple heavy/light-degree split with threshold $\sqrt{m}$\footnote{Indeed, there can be at most \smash{$\sum_{v : \deg(v)\le \sqrt{m}}\deg(v)^2 \le \sqrt{m}\sum_v \deg(v) \le 2m^{3/2}$} triangles involving a vertex of degree at most $\sqrt{m}$. On the other hand, there can be at most $O(\sqrt{m})$ vertices of degree at least $\sqrt{m}$, which form only $O(m^{3/2})$ triples.}.
    Therefore, if $P$ denotes the set of popular edges, then $n^3/K \le O(|P|^{3/2})$, which implies $|P| \ge \Omega(n^2/K^{2/3})$.
    Thus, the number of iterations is at most $I \le O(K^{2/3})$. With this bound, the running time bound simplifies to
    \begin{equation*}
        \parens*{ n^2(KL)^{O(1)} + \frac{n^{3+o(1)}}{K^{\Omega(\epsilon)}} + \frac{n^{3+o(1)}K^{2/3}}{L^{\Omega(\epsilon)}} }^{1+o(1)}.
    \end{equation*}
    By choosing $L = K^{O(1/\epsilon)}$ large enough, we can ensure that the second term dominates the third. The running time becomes
    \begin{equation*}
        \parens*{ n^2K^{O(1/\epsilon)} + \frac{n^{3+o(1)}}{K^{\Omega(\epsilon)}} }^{1+o(1)}.
    \end{equation*}
    For a small enough choice of $K = n^{\Theta(\epsilon)}$, this becomes $n^{3-\Omega(\epsilon^2)}$, which is truly subcubic.
\end{proof}
\section{Real-to-Integer Reduction for APSP}

For Real APSP, at first glance it seems hard to generalize the approach from \Cref{sec:exact-triangle} to negative triangles since the threshold choice $\Delta$ and popularity only bound the number of triangles summing to 0 up to $\pm \Delta$; the number of negative triangles is unbounded.
However, there is a simple fix that allows us to reuse basically the full previous reduction.
Namely, if there exists a \emph{very negative} triangle, that is, a triple $(i,k,j)$ satisfying 
\begin{equation*}
    A[i,k] + B[k,j] + C[i,j] < -\Delta,
\end{equation*}
in an instance of sufficiently low approximate rank, then a negative triangle is simple to detect.
Indeed, we can afford to iterate over the inner dimension of the approximate-rank decomposition and can then find a negative triangle using Fredman's trick and dominance product in truly subcubic time.
The next lemma formalizes this idea.
This step allows us to assume in the following that every negative triangle must also be a $\Delta$-approximate exact triangle.
Therefore, all other steps from \Cref{sec:exact-triangle} apply.

\begin{lemma}[Detecting Very Negative Triangles In Low-Rank Instances]
    \label{lem:detect-very-negative}
    Assume that Integer APSP is in time $O(n^{3-\epsilon}(\log u)^{O(1)})$.
    Let $\Delta \in \Real_{\geq 0}$, and let $(A,B,C)$ be a Real Negative Triangle instance, along with a $\Delta$-approximate rank-$r$ decomposition of~$C$. There is an algorithm running in time $\tilde O(rn^{3-\epsilon})$ that either outputs a negative triangle in $(A,B,C)$ or reports that every triangle has weight at least $-\Delta$.
\end{lemma}
\begin{proof}
    Take the given $\Delta$-approximate rank-$r$ decomposition $(U,V,S)$ of $C$, and use \Cref{obs:change-delta} to turn it into a $\Delta/2$-approximate rank-$5r$-decomposition.
    For all $\ell \in [5r]$, define
    \begin{align*}
        A_\ell[i,k] &\coloneqq \begin{cases}\mathmakebox[4cm][l]{A[i,k] + U[i,\ell]} & \text{if $A[i,k] \ne \bot$,}\\ \bot &\text{otherwise;} \end{cases}\\
        B_\ell[k,j] &\coloneqq \begin{cases}\mathmakebox[4cm][l]{-B[k,j] - V[\ell,j] - \Delta/2}& \text{if $B[k,j] \ne \bot$,}\\ \bot &\text{otherwise.} \end{cases}
    \end{align*}
    
    We aim to compute a \emph{dominance product} defined by
    \begin{equation*}
        D_\ell[i,j] \coloneqq \begin{cases}1 &\text{if $\exists k \in [n] : A_\ell[i,k] < B_\ell[k,j]$,}\\ 0 & \text{otherwise.}\end{cases} 
    \end{equation*}
    Once we have computed $D_\ell$, we check if there are $(i,j)$ such that $S[i,j] = \ell$ and $D_\ell[i,j] =1$, and find a witness $k$ for the dominance product in linear time. 
    Then, we output the triangle $(i,k,j)$ and stop the algorithm. If no witness is found for any $\ell$, we report that every triangle has weight at least $-\Delta$.

    To compute $D_\ell$, we use the fast algorithm for Integer APSP:
    First, since computing $D_\ell$ by definition requires only comparisons of matrix entries, we can simply replace all real entries by integers (by sorting all entries in $A_\ell$ and $B_\ell$ and assigning integers in a way that keeps the order).
    Then, negate the entries in $B_\ell$ to compute the integer \emph{min-plus product} 
    \begin{equation*}
        M_\ell[i,j] \coloneqq \min_{k \in [n]} A_\ell[i,k] - B_\ell[k,j].
    \end{equation*}
    Finally, the dominance product outputs can be inferred by setting $D_\ell[i,j] = 1$ if and only if $M_\ell[i,j] < 0$. (Of course, we may also compute each dominance product unconditionally in time $O(n^{(3+\omega)/2})$ using Matou{\v s}ek's algorithm~\cite{Matousek91}; however, we prefer our conditional statement since this leads to a combinatorial reduction.)

    \paragraph{Correctness.}
    Suppose there is a triangle $(i,k,j)$ with weight less than $-\Delta$ and let $\ell \coloneqq S[i,j]$.
    Then 
    \begin{equation*}
        -\Delta > A[i,k] + B[k,j] + C[i,j] \ge A[i,k] + B[k,j] + U[i,\ell] + V[\ell,j] - \Delta/2 = A_\ell[i,k] - B_\ell[k,j] - \Delta. 
    \end{equation*}
    So $A_\ell[i,k] < B_\ell[k,j]$ and the triangle $(i,k,j)$ will be detected in the dominance product.
    On the other hand, for any $(i,k,j)$ with $S[i,j] = \ell$ and $A_\ell[i,k] < B_\ell[k,j]$, we have
    \begin{equation*}
        0 > A_\ell[i,k] - B_\ell[k,j] \ge A[i,k] + B[k,j] + C[i,j]
    \end{equation*}
    and thus $A[i,k] + B[k,j] + C[i,j] < 0$. Therefore, if the algorithm outputs $(i,k,j)$, it must be negative.

    \paragraph{Running Time.}
    There are $5r$ iterations. In each iteration we compute a single dominance product via a single invocation of the fast Integer APSP algorithm, in time $\tilde O(n^{3-\epsilon})$. The total running time hence becomes $\tilde O(r n^{3-\epsilon})$, as claimed.
\end{proof}

Equipped with the previous lemma and everything from \Cref{sec:exact-triangle,sec:toolbox}, we can immediately prove \Cref{thm:apsp}.

\begin{proof}[Proof of \Cref{thm:apsp}]
    Suppose we are given an instance of Real Negative Triangle $(A,B,C)$ and 
    suppose there is an algorithm for Integer APSP and hence Integer Negative Triangle that runs in time $O(n^{3-\epsilon} (\log u)^{O(1)})$.
    We proceed in iterations in the same way as in the proof of \Cref{thm:exact-triangle}. First invoke \Cref{lem:threshold}, and suppose that it returns a scale $\Delta$ (we discuss the exceptional outcomes later). Then compute $P\sqcup U$ with \Cref{lem:decomp-pop}. 
    
    Let $C_P$ retain only the popular entries of $C$. On $(A,B,C_P)$, compute a $2\Delta$-approximate rank-$r = O(L\log n)$ decomposition of $C_P$ using \Cref{lem:popular-low-rank}, and refine it to scale $\Delta/2$ using \Cref{obs:change-delta}. The important difference comes here: Before listing all $\Delta$-approximate exact triangles with at least one edge in $P$, we first invoke \Cref{lem:detect-very-negative} at scale $\Delta/2$. Do the same with the roles of the three matrices exchanged to cover every possible position of a popular edge. If a call reports a negative triangle, we are done. Otherwise, every triangle with at least one popular edge has weight at least $-\Delta/2$. It suffices to list the $\Delta/2$-approximate exact triangles with a popular edge and check their original weights.
    The remaining proof is the same as for Exact Triangle; we discuss the remaining steps below.

    \paragraph{Only Popular Edges.}
    For solutions involving only popular edges, our approach is the same as in \Cref{lem:type1}. Restrict all matrices to popular entries by setting all other entries to $\bot$. By the choice of $\Delta$ (\Cref{lem:threshold}), the number of $\Delta$-approximate exact triangles is bounded by $T = O(n^3/K)$. Since, in addition, we detected very negative triangles with \Cref{lem:detect-very-negative}, the number of triangles $(i,k,j)$ with $A[i,k] + B[k,j] + C[i,j] \le \Delta$ is bounded by $T$.
    Whenever a reduction reports exceptional triangles or a listing algorithm returns triangles, we check their original real weights and stop if one is negative.
    
    We apply the reduction to Integer Exact Triangle listing (\Cref{lem:unified-popular-case}) at scale $\Delta/2$ with some parameter~$s$.
    By property (ii), every negative triangle in the remaining real instance is in $\mathcal T$ or becomes an integer exact triangle, since its weight lies in $[-\Delta/2,0)$.
    Focus on one resulting integer instance $(X,Y,Z)$.
    By \Cref{lem:unified-popular-case}, property (iii), any triangle $(i,k,j)$ in the integer instance with weight
    \begin{equation*}
        X[i,k] + Y[k,j] + Z[i,j] < 0
    \end{equation*}
    must also satisfy
    \begin{equation*}
        A[i,k] + B[k,j] + C[i,j] \le \Delta
    \end{equation*}
    in the real instance.
    Therefore, there are at most $T$ negative triangles in the integer instance. By \Cref{lem:unified-popular-case}, property (i), there are also at most $T$ zero-weight triangles. Thus, there are at most $2T = O(n^3/K)$ nonpositive integer triangles.
    Finally, we can list all such triangles with the listing-to-decision reduction for Integer Negative Triangle (\Cref{lem:negative-triangle-listing}), after subtracting $1$ from all entries in, say, $X$.

    Every step runs in exactly the same time as for Exact Triangle in \Cref{lem:type1}, so the running time analysis is identical.

    \paragraph{Popular and Unpopular Edges.}
    Now, we consider triangles with at least one popular edge and at least one unpopular edge, as in \Cref{lem:type2}.
    Suppose without loss of generality that we look for triangles where the $C$-entry is popular, and restrict $C$ accordingly. As in \Cref{lem:type2}, run the following steps separately on $(A_U,B,C)$ and $(A,B_U,C)$, where $A_U$ and $B_U$ retain only unpopular entries, so that every triangle in either instance has an unpopular edge.
    Both instances inherit the previously computed $\Delta/2$-approximate rank-$O(L\log n)$ decomposition of $C$.
    By unpopularity (\Cref{obs:number-solutions}), the number of $2\Delta$-approximate exact triangles with at least one unpopular entry is bounded by $T = O(n^3/L)$.
    By our additional detection of very negative triangles with \Cref{lem:detect-very-negative}, the number of triangles $(i,k,j)$ with $A[i,k] + B[k,j] + C[i,j] \le \Delta$ is hence also bounded by $T$.
    
    Now, we again apply the reduction to Integer Exact Triangle listing (\Cref{lem:unified-popular-case}) at scale $\Delta/2$ with some parameter $s$ and focus on one resulting integer instance $(X,Y,Z)$.
    Like in the previous case, we can use \Cref{lem:unified-popular-case}, properties (i) and (iii), to bound the number of triangles $(i,k,j)$ satisfying
    \begin{equation*}
        X[i,k] + Y[k,j] + Z[i,j] \le 0
    \end{equation*}
    by $O(n^3 / L)$.
    Finally, we list all such triangles with the listing-to-decision reduction for Integer Negative Triangle (\Cref{lem:negative-triangle-listing}), after subtracting $1$ from all entries in, say, $X$.

    Every step runs in exactly the same time as for Exact Triangle in \Cref{lem:type2}, so the running time analysis is identical.

    \paragraph{Iteration and Running Time.}
    The iterating step, which is discussed in \Cref{sec:only-unpop}, stays basically unchanged from Exact Triangle. Thus the same bound $|P|\ge\Omega(n^2/K^{2/3})$ applies, and these rounds also fit within $O(K^{2/3})$ iterations. Once \Cref{lem:threshold} reports that there are at most $O(n^3/K)$ $\infty$-approximate solutions, we list all such solutions with the listing-to-decision reduction (\Cref{lem:negative-triangle-listing}) by first replacing non-$\bot$ entries with, say, $-1$.
    
    The only remaining case is when \Cref{lem:threshold} ever reports that there are $\Omega(n^3 / K)$ exact solutions in the original instance. In this event, we proceed as before, but with $\Delta = 0$. This choice only affects the all-popular case, where we can no longer upper bound the number of $\Delta$-approximate triangles by $O(n^3 / K)$; however, the all-popular case can now be solved by one call to \Cref{lem:detect-very-negative} (since for $\Delta = 0$ every negative triangle is considered ``very negative'').

    The running time analysis is almost identical; the only difference is the additional running time $\tilde O(Ln^{3-\epsilon})$ caused by \Cref{lem:detect-very-negative}.
    Since we have to choose different values for $K$ and $L$, let us perform the calculations. Similarly to the proof of \Cref{thm:exact-triangle}, the running time is bounded by 
    \begin{equation*}
        \parens*{ n^2(KL)^{O(1)} + \frac{n^{3+o(1)}}{K^{\Omega(\epsilon)}} + \frac{n^{3+o(1)}K^{2/3}}{L^{\Omega(\epsilon)}} + K^{2/3}Ln^{3-\epsilon} }^{1+o(1)}.
    \end{equation*}
    By choosing $L = K^{O(1/\epsilon)}$ large enough, we can ensure that the second term dominates the third. The running time becomes
    \begin{equation*}
        \parens*{ n^2K^{O(1/\epsilon)} + \frac{n^{3+o(1)}}{K^{\Omega(\epsilon)}} + K^{O(1/\epsilon)}n^{3-\epsilon} }^{1+o(1)}.
    \end{equation*}
    For a small enough choice of $K = n^{\Theta(\epsilon^2)}$, this becomes $n^{3-\Omega(\epsilon^3)}$, which is truly subcubic.
\end{proof}

\section{Doubling Reduction for Real APSP}
\label{sec:proof-low-rank-to-small-doubling}

In this section, we prove the edge weight reduction (\Cref{lem:low-rank-to-small-doubling}), which allows us to transform any approximately low-rank matrix into an approximately uniform and low-doubling matrix with a potential-adjusting reduction.
Except for the $\Delta$-approximate part, the same chain of reductions was already carried out by Fischer~\cite{Fischer26} for the purpose of a universe reduction for APSP.
Many intermediate steps are in turn based on a chain of reductions for APSP with a truly sublinear number of distinct edge weights per node~\cite{AbboudFJVX25}.
We mostly have to follow the same reductions and proofs as in~\cite{Fischer26} in the approximate real setting.
For our use cases, some steps can be slightly shortened or presented in a slightly less general way; whenever possible, we take the opportunity to do so.
However, there are no big conceptual differences.

\subsection{Low-Rank to Slice-Uniform}

Let $\Delta \in \Realnneg$ and $d \in [n]$.
We call a matrix $A \in (\Real \cup \set{\bot})^{n \times n}$ \emph{$(d, \Delta)$-slice-uniform} if the non-$\bot$ entries of either each row or each column in $A$ have a $\Delta$-cover of size at most $d$. 
An instance $(A,B,C)$ is called $(d, \Delta)$-slice-uniform if at least one of $A$, $B$, and $C$ is $(d, \Delta)$-slice-uniform.

\begin{lemma}[Reduction from Approximately Low-Rank to Approximately Slice-Uniform Instances]\label{lem:low-rank-to-slice-uniform}
    For any $t \ge 1$, there is a potential-adjusting reduction from a Real Exact Triangle $\Delta$-approximate rank-$r$ instance $(A,B,C)$ to $O(t^2 \log r)$ $(r,\Delta)$-slice-uniform instances.
    It runs in time \[\tilde O\parens*{\frac{n^3}{t} + n^2rt^2}.\]
\end{lemma}
\begin{proof}
    Assume we are given a $\Delta$-approximate rank-$r$ decomposition $(U,V,S)$ of the matrix $C$, otherwise simply rotate the instance.
    Furthermore, we may assume that for some $R = O(\log n)$, and for all $\ell \in [r]$ and $i \in [n]$, we have
    \begin{equation*}
        |\set{j \in [n] : S[i,j] = \ell}| \le R \cdot \frac{n}{r},
    \end{equation*}
    that is, each inner dimension of the rank decomposition is selected the same number of times per row up to factors of $O(\log n)$ (see~\cite[Lemmas 3.5 and 4.4]{Fischer26} for the corresponding unchanged arguments of this assumption).

    Let $\mathcal{T} \gets \emptyset$ be the set of exceptional triples, and let $\mathcal{I} \gets \emptyset$ be the set of $(r,\Delta)$-slice-uniform potential adjustments of $(A,B,C)$.
    Consider the algorithm that enumerates $\ell \gets 1, \ldots, r$. In iteration $\ell$, we ensure that each $\Delta$-approximate exact triangle $(i,k,j)$ with $S[i,j] = \ell$ is either inserted into $\mathcal T$ or still present in an instance in $\mathcal I$.

    \begin{enumerate}
        \item \emph{(Heavy/Light Setup)}
            Define matrices $A_\ell$ and $B_\ell$ of size $n \times n$ by
            \begin{align*}
                A_\ell[i,k] &\coloneqq  A[i,k] + U[i,\ell],\\
                B_\ell[k,j] &\coloneqq -(B[k,j] + V[\ell, j]).
            \end{align*}
            Note that this way for all pairs $(i,j)$ with $S[i,j] = \ell$, the condition $A[i,k] + B[k,j] + C[i,j] = 0 \pm \Delta$ implies $A_\ell[i,k] = B_\ell[k,j] \pm 2\Delta$.
            For the remainder of this algorithm, we classify the non-$\bot$ entries of $A_\ell$ as either \emph{light} or \emph{heavy}.
            We call a non-$\bot$ entry $(i,k)$ light if
            \begin{equation*}
                \abs{\set{i' \in [n] : A_\ell[i,k] = A_\ell[i',k] \pm 4\Delta}} \le \frac{n}{rt}
            \end{equation*}
            and heavy otherwise.

            Creating $A_\ell$ and $B_\ell$ and classifying entries as either heavy or light takes time $\tilde O(n^2r)$ across all $r$ iterations.

        \item \emph{(Light Edges)}
            We enumerate all triples $(i,k,j)$ with $A_\ell[i,k] = B_\ell[k,j] \pm 2\Delta$ for which $(i,k)$ is light as follows.
            Enumerate all pairs $(k,j) \in [n]^2$.
            If there exists a light edge $(i',k)$ with $A_\ell[i',k] = B_\ell[k,j] \pm 2\Delta$, enumerate all rows $i$ with $A_\ell[i,k] = B_\ell[k,j]\pm 2\Delta$. There are at most $n/(rt)$ such rows, since each satisfies $A_\ell[i,k] = A_\ell[i',k]\pm 4\Delta$ and $(i',k)$ is light.
            We insert each such triple into $\mathcal T$.

            This step correctly deals with all $\Delta$-approximate solutions involving light edges in time $\tilde O(n^3/t+n^2r)$ across all $r$ iterations.

        \item \emph{(Heavy Edges)}
            We distinguish two cases depending on the number of heavy edges:
            
            \begin{enumerate}
                \item \emph{(Few Heavy Edges)}
                    If there are at most $n^2/t$ heavy edges, we can enumerate all relevant triples by brute force as follows.
                    Enumerate all heavy pairs $(i,k) \in [n]^2$, and enumerate all columns $j \in [n]$ with $S[i,j] = \ell$.
                    Each enumerated triple $(i,k,j)$ that forms a $\Delta$-approximate exact triangle is inserted into $\mathcal T$.
                    
                    This clearly takes care of all heavy triples.
                    By the initial regularity assumptions of the rank decomposition, for each $i \in [n]$, there are at most $\tilde O(n/r)$ columns $j$ with $S[i,j] = \ell$.
                    Hence, the running time of this step is $\tilde O(n^2/t \cdot n/r)$ and thus $\tilde O(n^3/t)$ across all iterations.

                \item \emph{(Many Heavy Edges)}
                    The final case is that there are more than $n^2 / t$ heavy pairs.
                    Let $A_H$ be the matrix $A$ restricted to the heavy pairs (all other entries are set to $\bot$).
                    We take care of the full instance $(A_H, B, C)$ using potential-adjustments and afterwards globally remove all heavy entries from $A$.
                    Construct matrices $A'$ and $C'$ by
                    \begin{align*}
                        A'[i,k] &\coloneqq \begin{cases}A[i,k] + U[i,\ell], &\text{if $(i,k)$ is heavy,}\\\bot, &\text{otherwise;} \end{cases}\\
                        C'[i,j] &\coloneqq C[i,j] - U[i,\ell].
                    \end{align*}
                    By definition, $(A', B, C')$ is a potential-adjusting restriction of $(A,B,C)$ and retains all heavy exact triangle solutions.
                    Furthermore, $A'$ is $(4rt,\Delta)$-slice-uniform. To see this, in each column greedily select the leftmost uncovered heavy value $a$ and cover all heavy values in $[a,a+8\Delta]$ by four intervals of width $2\Delta$. The selected values are more than $8\Delta$ apart, so their neighborhoods $[a-4\Delta,a+4\Delta]$ are disjoint. Each neighborhood contains more than $n/(rt)$ entries of $A_\ell$, hence there are at most~$rt$ selections and at most $4rt$ cover intervals.
                    Then, we trivially reduce $(A',B,C')$ to $4t$ instances that are $(r,\Delta)$-slice-uniform and insert these instances in $\mathcal I$.

                    This way we have dealt with all triangles $(i,k,j)$ where $(i,k)$ is heavy. Therefore, we can safely assign all heavy entries $A[i,k]$ to $\bot$ for all future iterations.
                    Regarding the running time, constructing the slice-uniform instances takes near-linear time in the output $\tilde O(n^2t)$ and adds $O(t)$ instances to $\mathcal I$. Since there are at least $n^2/t$ heavy edges and we remove them globally after the iteration, we execute this step at most $t$ times in total. Therefore, across all iterations, the running time of this step is bounded by $\tilde O(n^2t^2)$ and $\abs{\mathcal I} \le 4t^2$.
            \end{enumerate}
    \end{enumerate}
    The correctness and running time of the reduction are already analyzed per step. In total, the running time is bounded by $\tilde O(n^3/t + n^2rt^2)$, as claimed. With the initial regularity assumption, the number of created instances is bounded by $O(t^2 \log r)$.
\end{proof}

\subsection{Slice-Uniform to Uniform}

Call an instance $(A,B,C)$ \emph{$(D, \Delta)$-uniform} if the set $X$ of all non-$\bot$ entries of $A$, $B$, and $C$ satisfies $\cabs X \le D$.

\begin{lemma}[Reduction from Approximately Slice-Uniform to Approximately Uniform Instances]\label{lem:slice-uniform-to-uniform}
    For any $t \ge 1$, there is a potential-adjusting reduction from a  Real Exact Triangle instance in which one matrix is $(d, \Delta)$-slice-uniform to $O(t^{41}\log^2 n)$ instances which are $(d, \Delta)$-uniform. 
    The reduction runs in time \[\tilde O\parens*{n^2dt^{41} + \frac{n^3}{t} }.\]
\end{lemma}

Before we prove \Cref{lem:slice-uniform-to-uniform}, we establish two convenient lemmas. Denote by $P^{\pm\Delta}_s(A, B) \subseteq A+B$ the set of \emph{approximately $s$-popular sums}, consisting of elements $x \in A+B$ with at least $s$ pairs $(a,b) \in A\times B$ that satisfy $x = a+b \pm \Delta$.

\begin{lemma}[Approximating Approximately Popular Sums] \label{lem:pop-sum-approx}
    Given sets $X,Y \subseteq \Real$, $s > 0$, and $\Delta \in \Realnneg$, there is an algorithm that computes a set $P \subseteq \Real$ such that:
    \begin{enumerate}[label=(\roman*)]
        \item $P$ is a $\Delta$-cover for $P^{\pm\Delta}_{2s}(X,Y)$,
        \item for all $p \in P$, we have $\abs{\set{(x,y) \in X \times Y : p = x+y \pm \Delta}} \ge s$,
        \item for all $z \in \Real$, there are at most two $p \in P$ with $p = z \pm \Delta$.
    \end{enumerate}
    The algorithm runs in randomized time $\tilde O(|X||Y|/s)$.
\end{lemma}
\begin{proof}
    We first sample a multiset $S$ made up of $R\abs{X}\abs{Y}/s$ independent uniformly random elements from $X \times Y$ for some $R = \Theta(\log n)$.
    Define the candidate set $Q\coloneqq\set{x + y : (x, y) \in S} + \set{-\Delta, 0, \Delta}$. Let $P \subseteq Q$ consist of all centers $p$ for which at least $1.5R$ pairs $(x', y') \in S$ satisfy $p = x' + y' \pm\Delta$. Next, scan $P$ in increasing order. Remove a center $p$ if its current predecessor $p'$ and successor $p''$ both exist and satisfy $p''-p' \le 2\Delta$. Finally, we return $P$.

    \paragraph{Correctness.}
    The final deletion step clearly guarantees property (iii). At the same time, we only delete elements which cannot have been necessary for property~(i). We thus continue proving properties~(i) and~(ii) pretending that we did not delete any elements from $P$.

    For property (i), consider any sum $z \in P_{2s}^{\pm \Delta}(X,Y)$. There are at least $2s$ pairs $(x,y) \in X \times Y$ with $x + y = z \pm \Delta$.
    By Chernoff's bound, we have sampled at least $1.5R$ such pairs in $S$ with high probability. 
    Let $a$ and $b$ be the smallest and largest sampled sums in $[z-\Delta,z+\Delta]$. Both $b-\Delta$ and $a+\Delta$ belong to $Q$, and their width-$2\Delta$ intervals contain all these sampled sums, so both enter $P$. Since $z\in[b-\Delta,a+\Delta]$ and this interval has length at most $2\Delta$, at least one of these centers is within $\Delta$ of $z$. A union bound over at most $|X||Y|$ possible sums $z$ proves property (i).
    
    Similarly, consider any $z \in X + Y + \set{-\Delta,0,\Delta}$ for which there are at most $s$ choices of $(x,y) \in X\times Y$ with $x + y = z \pm \Delta$.
    Chernoff's bound again shows that there will be at most $1.1R$ such pairs sampled in $S$ with high probability.
    Therefore, conditioned on this event, $z$ itself is not in $P$. Since $P\subseteq Q\subseteq X + Y + \set{-\Delta,0,\Delta}$, a union bound over this fixed set of at most $3|X||Y|$ possible centers proves property (ii).

    \paragraph{Running Time.}
    The multiset $S$ has size $\tilde O(\abs{X}\abs{Y}/s)$ and $|Q|\le3|S|$. Sorting the sampled sums and the candidates allows us to compute all sample counts by a linear sweep. The final deletions take one linear scan over the sorted centers. The running time is $\tilde O(\abs{X}\abs{Y}/s)$.
\end{proof}

\begin{lemma}[Popular Sum Decomposition] \label{lem:pop-sum-decomp}
    Let $\Delta \in \Realnneg$, let $X_1, \dots, X_n, Y_1, \dots, Y_n \subseteq \Real$ be $\Omega(\Delta)$-spaced sets of size at most $d$, and let $p \geq 1$. Then there exist an integer $m=O(p^2)$ and partitions
    \begin{align*}
        X_i &= X_i^* \sqcup \bigsqcup_{g=1}^{m} X_{i, g}, \\
        Y_j &= Y_j^* \sqcup \bigsqcup_{h=1}^{m} Y_{j, h}
    \end{align*}
    satisfying the following two properties:
    \begin{enumerate}[label=(\roman*)]
        \item
            There are sets $S_g \subseteq \Real$ of size $|S_g| \leq d$ and shifts $s_{i, g} \in \Real$ such that $S_g + s_{i, g}$ is a $\Delta$-cover for $X_{i,g}$ (for all $i \in [n], g \in [m]$), and\newline
            there are sets $T_h \subseteq \Real$ of size $|T_h| \leq d$ and shifts $t_{j, h} \in \Real$ such that $T_h + t_{j, h}$ is a $\Delta$-cover for  $Y_{j, h}$ (for all $j \in [n], h \in [m]$).
        
        \item
            $\abs{\set{(i, j) \in [n]^2 : P^{\pm\Delta}_{2d / p}(X_i^*, Y_j) \neq \emptyset}} \leq n^2 / p$, and\newline
            $\abs{\set{(i, j) \in [n]^2 : P^{\pm\Delta}_{2d / p}(X_i, Y_j^*) \neq \emptyset}} \leq n^2 / p$.
    \end{enumerate}
    Moreover, there is an algorithm that, given the sets $X_i, Y_j$, computes the partitions along with the sets $S_g, T_h$ and the shifts $s_{i, g}, t_{j, h}$ in randomized time \smash{$\tilde O(n^2dp^3)$}.
\end{lemma}
\begin{proof}
    We show how to construct the partitions of the sets $X_i$ and the $S_g$ and $s_{i,g}$; the other part is symmetric. Fix a constant $c_0\ge1$ such that all $X_i$ and $Y_j$ are $\Delta/c_0$-spaced. Every interval of width $2\Delta$ then contains at most $2c_0+1$ elements of any one of these sets. Consider the algorithm that runs the following steps for $g = 1, 2, \dots$:
    \begin{enumerate}
        \item Apply \Cref{lem:pop-sum-approx} with parameter $d/p$. For each $(i,j) \in [n]^2$, it computes a set $P_{i,j} \subseteq \Real$ that is a $\Delta$-cover of $P_{2d/p}^{\pm \Delta}(X_i, Y_j)$ and each $p \in P_{i,j}$ covers sums from at least $d/p$ pairs in $X_i\times Y_j$.
        
        \item If $\abs{\set{(i,j) \in [n]^2 : P_{i,j} \ne \emptyset}} \le n^2/p$, pick $X_i^* \coloneqq X_i$ for all $i \in [n]$ and stop the algorithm.
    
        \item Otherwise, pick some $j \in [n]$ for which there are at least $n/p$ many $i \in [n]$ such that $P_{i,j} \ne \emptyset$. 
        Choose $S_g \coloneqq -Y_j$.
        We call an index $i$ \emph{good} if $P_{i,j} \ne \emptyset$ and \emph{bad} otherwise.
        For each bad $i$, choose $X_{i,g} \coloneqq \emptyset$ and $s_{i,g}$ arbitrarily.
        For each good $i$, instead pick an arbitrary cover center $s_{i,g} \in P_{i,j}$.
        Then, pick \[X_{i,g} \coloneqq X_i \cap \bigcup_{s \in S_g} [s + s_{i,g} - \Delta, s+s_{i,g} + \Delta],\] that is, as the set of elements in $X_i$ that are covered by a number in $S_g + s_{i,g}$.
        Afterwards, remove $X_{i,g}$ from $X_i$ and continue with the next iteration.
    \end{enumerate}
    
    \paragraph{Correctness.}
    All properties are trivially satisfied after the algorithm terminates except for $g=O(c_0p^2)=O(p^2)$, which we show in the following. 
    Consider iteration $g$ and suppose we chose index $j$. Let $i$ be a good index.
    We claim that $X_{i,g}$ has size at least $d/((2c_0+1)p)$. For each fixed $x\in X_i$, the spacing of $Y_j$ allows at most $2c_0+1$ elements $y\in Y_j$ with $x+y=s_{i,g}\pm\Delta$. Thus each covered element of $X_i$ accounts for at most $2c_0+1$ of the representing pairs.
    Indeed by the choice of $s_{i,g} \in P_{i,j}$, we have that 
    \begin{equation*}
        \abs{X_{i,g}} = \big| X_i \cap \bigcup_{s \in S_g} [s + s_{i,g} - \Delta, s+s_{i,g} + \Delta]\big| \ge \frac{\abs{\set{(x,y)\in X_i\times Y_j : x+y=s_{i,g}\pm\Delta}}}{2c_0+1} \ge \frac{d}{(2c_0+1)p}.
    \end{equation*}

    Thus, the size of each good set $X_i$ is decreased by $\Omega(d/(c_0p))$ and there are at least $n/p$ good indices. So we reduce $\sum_{i \in [n]} \abs{X_i}$ by $\Omega(nd/(c_0p^2))$ in each step, while initially we have $\sum_{i \in [n]} \abs{X_i} \le nd$. The bound on the number of iterations follows.
    
    \paragraph{Running Time.}
    In a single iteration, we run \Cref{lem:pop-sum-approx} a total of $n^2$ times with a running time of $\tilde O(d^2/(d/p)) = \tilde O(dp)$ each. Step 2 takes time $O(n^2)$. Step 3 is simple to implement in time $O(n^2d)$. With $O(c_0p^2)$ iterations in total, the running time is $\tilde O(c_0n^2dp^3)=\tilde O(n^2dp^3)$, recalling that $c_0$ is fixed.
\end{proof}

\begin{proof}[Proof of \Cref{lem:slice-uniform-to-uniform}]
    By rotating or transposing, we can assume without loss of generality that $A$ is $(d,\Delta)$-slice-uniform and the set of entries $X$ of each \emph{row} of $A$ satisfies $\cabs X \le d$.

    \paragraph{Outer Regularization}
    We start with a preprocessing step that allows us to assume afterwards that each element in the $\Delta$-cover covers roughly the same number of elements in each row of $A$ and column of $B$.
    To this end, we partition $A$ into $L = \lceil \log n \rceil$ matrices $A_0, \ldots, A_L$ where we include in the matrix $A_\ell$ all entries $A[i,k]$ for which there are between $2^\ell$ and $2^{\ell + 1}$ entries $A[i,k']$ in its row with $A[i,k'] = A[i,k] \pm \Delta$. Similarly, we partition $B$ into $B_0, \ldots, B_L$ according to the number of approximately equal column entries. We are left with $(L+1)^2$ instances $(A_\ell, B_{\ell'}, C)$ for $0 \le \ell, \ell' \le L$.
    
    For the remainder of the proof, fix one such instance and write $A = A_\ell$ and $B = B_{\ell'}$ for simplicity of notation.
    For each $i \in [n]$, compute the greedy minimum $\Delta$-cover $X_i$ for the $i$-th row of $A$ and also compute the greedy minimum $\Delta$-cover $Y_j$ for the entries of the $j$-th column of $B$ for each $j \in [n]$.
    Let $d_A \coloneqq \min\set{d, n/2^\ell}$ and $d_B \coloneqq n/2^{\ell'}$ and note that for each row $i$ of $A$, we have that $|X_i| \le d_A$. The same claim holds for columns of $B$ with $|Y_j| \le d_B$. 
    We finally argue that we may also assume that $d_B \le d_A\cdot t$ in the following. Indeed, if $d_B > d_A \cdot t$, we can enumerate all solutions in the instance efficiently as follows.
    Enumerate all pairs $(i,j) \in [n]^2$, enumerate all of the at most $d_A$ elements $x \in X_i$, and enumerate all $k \in [n]$ with $B[k,j] = - C[i,j] - x \pm 2\Delta$. Insert all enumerated triangles $(i,k,j)$ into $\mathcal T$.

    We take care of all $\Delta$-approximate exact triangles in this step if the assumptions are not met. Indeed, if $(i,k,j)$ is a $\Delta$-approximate exact triangle, then $B[k,j] = -C[i,j] - A[i,k] \pm \Delta$, and because $A[i,k] = x \pm \Delta$, we have also $B[k,j] = -C[i,j] - x \pm 2\Delta$. Therefore, this step is correct. For the running time, notice that there are only $|X_i| \le d_A$ many elements $x$ per row to enumerate. For each cover element, there are only constantly many elements $y \in Y_j$ to enumerate that would complete a $\Delta$-approximate exact triangle. Each such $y$ covers at most $O(n/d_B)$ values $B[k,j]$. The total time to enumerate everything is hence bounded by \[\tilde O\parens*{n^2 d_A\cdot \frac{n}{d_B}} = \tilde O \parens*{\frac{n^3}{t}}.\]

    \paragraph{Uniformization.}
    Recall that $X_i$ and $Y_j$ have size at most $d' \coloneqq \max\set{d_A, d_B}$. Run \Cref{lem:pop-sum-decomp} on them for some parameter $p \ge 1$ to be determined later.
    Let $m=O(p^2)$ denote the number of parts in each returned partition.
    We call a $\Delta$-approximate exact triangle $(i,k,j)$ \emph{$A$-exceptional} if $A[i,k]$ is covered by some $x \in X_i^*$, and \emph{$B$-exceptional} if $B[k,j]$ is covered by some $y \in Y_j^*$, and \emph{ordinary} otherwise. 
    
    The running time of this setup is bounded by the running time of \Cref{lem:pop-sum-decomp}, that is, $\tilde O(n^2d'p^3) = \tilde O(n^2dtp^3)$.

    \begin{enumerate}
        \item \emph{(Exceptional Triangles)}
            We will explicitly enumerate all exceptional $\Delta$-approximate exact triangles. We describe how to enumerate all $A$-exceptional $\Delta$-approximate exact triangles and omit the analogous proof for $B$-exceptional triangles (including exchanging the roles of $d_A$ and $d_B$).
            First, apply \Cref{lem:pop-sum-approx} for all $(i,j) \in [n]^2$ to obtain a set $P_{i,j}$ that is a $3\Delta$-cover for $P_{24d'/p}^{\pm 3\Delta}(X_i^*, Y_j)$ where each element in $P_{i,j}$ covers sums from at least $12d'/p$ pairs in~$X_i^*\times Y_j$ up to $\pm 3\Delta$ and each element is covered by at most 2 numbers in $P_{i,j}$.
            We then call such a triangle $(i,k,j)$ \emph{popular} if $-C[i,j] = z \pm 6\Delta$ for some $z \in P_{i,j}$ and \emph{unpopular} otherwise.
            We solve both cases separately.

        \begin{enumerate}
            \item \emph{(Popular Triangles)}
                In this case, we simply use that the number of pairs $(i,j)$ supporting popular exceptional triangles is bounded.
                If $P_{i,j}\ne\emptyset$, any one of its centers covers sums from at least~$12d'/p$ pairs in an interval of width $6\Delta$. Partition this interval into six intervals of width $\Delta$; one contains sums from at least $2d'/p$ pairs. Any sum in this subinterval belongs to \smash{$P_{2d'/p}^{\pm\Delta}(X_i^*,Y_j)$}.
                Therefore, we know by definition of $X^*_i$ (property (ii) of \Cref{lem:pop-sum-decomp}) that the number of nonempty $P_{i,j}$ is at most $n^2/p$. For each such pair, we enumerate all possible $k \in [n]$ and store the triangle $(i,k,j)$ in~$\mathcal T$.
                
                The running time of this step is clearly bounded by $O(n^3/p)$.
             
            \item \emph{(Unpopular Triangles)}
                We enumerate all pairs $(i,j) \in [n]^2$ where $-C[i,j]$ is not covered by $P_{i,j}$ up to $\pm 6 \Delta$.
                We enumerate all $x \in X_i^*$ and test if a suitable $y$ exists that satisfies $x + y = -C[i,j] \pm 3\Delta$. If it does, we list all $k$ with $A[i,k] = x \pm \Delta$ and insert $(i,k,j)$ into $\mathcal T$.

                For the running time, suppose that $A[i,k] + B[k,j] + C[i,j] = 0 \pm \Delta$ where $A[i,k] = x \pm \Delta$, $B[k,j] = y \pm \Delta$ for some $x \in X_i^*$ and $y \in Y_j$, and no $z \in P_{i,j}$ satisfies $-C[i,j] = z \pm 6\Delta$.
                This implies that $x + y = - C[i,j] \pm 3\Delta$.
                On the other hand, since $-C[i,j]$ is not covered by $P_{i,j}$ up to $\pm 6\Delta$, fewer than $48d'/p$ pairs $(x,y)\in X_i^*\times Y_j$ satisfy $x+y=-C[i,j]\pm3\Delta$. Otherwise, one half of this width-$6\Delta$ interval would contain sums from at least $24d'/p$ pairs; any one of those sums would belong to $P_{24d'/p}^{\pm3\Delta}(X_i^*,Y_j)$ and hence have a cover center in $P_{i,j}$ within $6\Delta$ of $-C[i,j]$.
                Notice that for each pair $(i,j)$ we list $d_A \le d$ values $x$ and in total at most $O(d'/p)$ many pairs $(x,y)$. For each such pair $(x,y)$ we enumerate at most $O(n/d_A)$ values of $k$. The time for this step is hence bounded by 
                \begin{equation*}
                    O\parens*{ n^2\parens*{ d + \frac{d'}{p}\cdot \frac{n}{d_A} } } =  O\parens*{ n^2\parens*{ d + \frac{nt}{p} } } =  O\parens*{ n^2d + \frac{n^3t}{p} }.
                \end{equation*}
        \end{enumerate}

        \item \emph{(Ordinary Triangles)}
            Enumerate all pairs $g, h \in [m]$. Define matrices $A_g$, $B_h$, and $C_{g,h}$ as 
            \begin{align*}
                A_g[i,k] &\coloneqq \begin{cases}\mathmakebox[2.8cm][l]{A[i,k] - s_{i,g}} &\text{if $A[i,k] = x \pm \Delta$ for some $x \in X_{i,g}$,}\\\bot &\text{otherwise;} \end{cases}\\
                B_h[k,j] &\coloneqq \begin{cases}\mathmakebox[2.8cm][l]{B[k,j] - t_{j,h}} &\text{if $B[k,j] = y \pm \Delta$ for some $y \in Y_{j,h}$,}\\\bot &\text{otherwise;} \end{cases}\\
                C_{g,h}[i,j] &\coloneqq \begin{cases}\mathmakebox[2.8cm][l]{C[i,j] + s_{i,g} + t_{j,h}} &\text{if $C[i,j] \ne \bot$,}\\\bot &\text{otherwise.} \end{cases}
            \end{align*}
            We can assume that each triangle appears in only one instance since each number is only covered by one element of a $\Delta$-cover (\Cref{obs:delta-cover}).
            Moreover, the value of the triangle is unchanged since all $s_{i,g}$ and $t_{j,h}$ cancel:
            \begin{equation*}
                A_g[i,k] + B_h[k,j] + C_{g,h}[i,j] = A[i,k] + B[k,j] + C[i,j].
            \end{equation*}

            Now, we can verify that $A_g$ and $B_h$ are $(d',2\Delta)$-uniform: The set $X_{A_g}$ of all non-$\bot$ entries in $A_g$ is covered by the $2\Delta$-cover $S_g$, by \Cref{lem:pop-sum-decomp}, property (i).
            Similarly, all non-$\bot$ entries in $B_h$ are covered by the $2\Delta$-cover $T_h$.
            Since $|S_g|, |T_h| \le d'$, each matrix has a $2\Delta$-cover of size at most $d'$ and hence a $\Delta$-cover of size at most $2d'$, by splitting each covering interval into two.

            Let $q\ge1$ be another parameter to be determined later. In the following, we modify $C_{g,h}$ to also become $(O(d'q),\Delta)$-uniform to make the whole instance $(A_g, B_h, C_{g,h})$ uniform. Define the set
            \begin{equation*}
                P \coloneqq \set{z \in \Real : \abs{\set{(x, y) \in S_g \times T_h : x + y + z = 0 \pm 5\Delta}} \ge d'/q}.
            \end{equation*}
            We can compute $P$ in time $\tilde O(|S_g| \, |T_h|) = \tilde O(d'^2)$, represented as a union of intervals. Each membership query in $P$ can then be answered in time $O(\log n)$.
            We call a pair $(i,j)$ \emph{popular} if $C_{g,h}[i,j] \in P$ and \emph{unpopular} otherwise. 
            We solve both cases separately.

            \begin{enumerate}
                \item \emph{(Popular Entries)}
                    Let $C_{g,h}^{\text{pop}}$ be the restriction of $C_{g,h}$ to the popular entries with all unpopular entries replaced by $\bot$. To bound its cover size, take a maximal subset $Q\subseteq P$ whose points are more than $10\Delta$ apart. The intervals of width $10\Delta$ around points of $Q$ are disjoint, and each contains negated sums from at least $d'/q$ pairs, so $|Q|\le |S_g||T_h|/(d'/q)$. By maximality, $Q$ is a $10\Delta$-cover of $P$; splitting each interval into two gives a $5\Delta$-cover.
                    Observe that 
                    \begin{equation*}
                        |P|_{\pm 5\Delta} \le O(|S_g||T_h|/(d'/q)) = O(d'q)
                    \end{equation*} 
                    and therefore $P$ also has a $\Delta$-cover of size $O(d'q)$. Hence, the instance \smash{$(A_g, B_h, C_{g,h}^{\text{pop}})$} is $(O(d'q),\Delta)$-uniform. We can trivially reduce this further to $O((d'q/d)^3)=O((tq)^3)$ instances that are $(d,\Delta)$-uniform, which we store in $\mathcal I$.

                    There are $m^2=O(p^4)$ possible choices for $(g,h)$, so the total running time of this step is $\tilde O(p^4t^3q^3n^2)$.

                \item \emph{(Unpopular Entries)}
                    In this case, we enumerate all unpopular pairs $(i,j)$. 
                    By the definition of~$P$, fewer than $d'/q$ pairs $(x, y) \in S_g \times T_h$ satisfy $x + y + C_{g, h}[i,j] = 0 \pm 5\Delta$. We enumerate all $x \in S_g$ and test whether there is $y \in T_h$ satisfying this condition. If there is, we list all $k$ with $A_g[i, k] = x \pm 2\Delta$ and insert $(i,k,j)$ into $\mathcal T$.

                    For correctness, every $\Delta$-approximate exact triangle $(i, k, j)$ in this instance has representatives $x \in S_g$ and $y \in T_h$ with $A_g[i, k] = x \pm 2\Delta$ and $B_h[k, j] = y \pm 2\Delta$. Together with $A_g[i, k] + B_h[k, j] + C_{g, h}[i, j] = 0 \pm\Delta$, this gives $x + y + C_{g, h}[i, j] = 0 \pm 5\Delta$, so we certainly list $k$ and thus include $(i, k, j)$ into $\mathcal T$.
                    
                    For the running time, notice that for each pair $(i,j)$, we list $d'$ values of $x$ and in total at most $O(d'/q)$ pairs $(x,y)$. For each such pair, we enumerate at most $O(n/d_A)$ values of $k$. The total time of this step is hence bounded by
                    \begin{equation*}
                        \tilde O\parens*{p^4n^2 \parens*{d' + \frac{d'}{q}\cdot \frac{n}{d_A}}} = \tilde O\parens*{p^4n^2 \parens*{dt + \frac{d_At}{q}\cdot \frac{n}{d_A}}} = \tilde O\parens*{n^2dtp^4  + \frac{n^3tp^4}{q} }.
                    \end{equation*}
                
            \end{enumerate}
        
    \end{enumerate}

    This completes the algorithm description and correctness arguments.
    
    \paragraph{Running Time.}
    The initial preprocessing runs in time $\tilde O(n^3/t)$ and creates only $O(\log^2 n)$ many instances. Every other contribution has been analyzed before and leads in total to a running time bound of 
    \begin{equation*}
        \tilde O\parens*{ n^2dt^3p^4q^3 + \frac{n^3}{t} + \frac{n^3t}{p} + \frac{n^3tp^4}{q} }.
    \end{equation*}
    Setting $p \coloneqq t^2$ and $q \coloneqq t^{10}$, the bound simplifies to $\tilde O(n^2dt^{41}+ n^3/t )$, as claimed. The number of created instances with this choice of parameters is bounded by $O(t^{41}\log^2n)$.
\end{proof}

\subsection{Uniform to Uniform Regular}
Call a matrix $A$ \emph{$(\rho, \Delta)$-regular} if there is no $x \in \Real$ such that there are more than $\rho n$ entries $A[i,k]$ with $A[i,k] = x \pm \Delta$ in the same row or column of $A$.
We say that an instance is $(\rho, \Delta)$-regular, if all three matrices are $(\rho, \Delta)$-regular.
In other words, if we consider a $\Delta$-cover $X$ of the entries in an arbitrary row or column of $A$, no element covers more than $\rho n$ many entries.

\begin{lemma}[Reduction from Approximately Low-Rank to Approximately Uniform and Regular Instances]\label{lem:low-rank-to-uniform-regular}
    For any $t \ge 1, \Delta \ge 0$, there is a potential-adjusting reduction from a Real Exact Triangle instance $(A,B,C)$ where one matrix has $\Delta$-approximate rank-$r$ to $t^{875}n^{o(1)}$ instances that are $(r, \Delta)$-uniform and $(1/r, \Delta)$-regular. The reduction runs in time \[\parens*{rt^{875}n^2 + \frac{n^3}{t}}^{1+o(1)}.\]
\end{lemma}
\begin{proof}
    We use the exact same proof as in \cite{Fischer26} modulo the approximate versions of rank, slice-uniform, uniform, and regular. 

    \begin{enumerate}
        \item \emph{(Low-Rank to Slice-Uniform)}
            Run the potential-adjusting reduction from \Cref{lem:low-rank-to-slice-uniform} with some parameter $t_\text{sl}$ to transform the low-rank instance into $(r,\Delta)$-slice-uniform instances. Denote the output by $(\mathcal I_\text{sl}, \mathcal T_\text{sl})$.
            
            This step runs in time $\tilde O(n^3/t_\text{sl} + n^2rt_\text{sl}^2)$.

        \item \emph{(Slice-Uniform to Uniform)}
            We further reduce each of the obtained instances $(A',B',C') \in \mathcal I_\text{sl}$ into $(r, \Delta)$-uniform instances $(\mathcal I', \mathcal T')$ by running \Cref{lem:slice-uniform-to-uniform} with some parameter $t_\text{unif}$. Let $\mathcal I_\text{unif}$ be the union of all $\mathcal I'$ and $\mathcal T_\text{unif}$ the union of all $\mathcal T'$.
            
            This reduction is still potential-adjusting and runs in time $\tilde O(\abs{\mathcal I_\text{sl}}\cdot (n^3/t_\text{unif} + n^2rt_\text{unif}^{41}))$.

        \item \emph{(Uniform to Uniform Regular)}
            We reduce the instances in $\mathcal I_\text{unif}$ further to regular instances $\mathcal I_\text{reg}$. Set initially $\mathcal I_\text{reg} \gets \emptyset$.
            In this reduction step we will miss some $\Delta$-approximate exact triangles, but we ensure that each such triangle is present in one of the instances $(A_\text{sm}, B, C)$, $(A, B_\text{sm}, C)$, or $(A,B,C_\text{sm})$ that we will construct.
            We will guarantee that these remaining instances have low approximate rank, so we will maintain low-$\Delta$-approximate-rank decompositions of $A_\text{sm}$, $B_\text{sm}$, and $C_\text{sm}$.
            Initially, we set every entry of these matrices equal to $\bot$ and start with a trivial $\Delta$-approximate rank-0 decomposition.

            Enumerate the instances $(A', B', C') \in \mathcal I_\text{unif}$. 
            Since they are obtained from potential-adjusting reductions, we also know potential adjustments $u$ and $v$ such that 
            \begin{equation*}
                A'[i,k] = A[i,k] + u[i] + v[k] \quad \text{or} \quad A'[i,k] = \bot
            \end{equation*}
            for all entries.
            Let $q$ be a parameter determined later.
            Let $X_i$ be the greedy minimum $\Delta$-cover of the entries of the $i$-th row of $A'$, with $|X_i|\le r$, and let $Y_k$ be the corresponding cover for the $k$-th column. In each cover the centers are more than $2\Delta$ apart, so every non-$\bot$ entry belongs to a unique cover interval, which we call its \emph{bucket}.
            We call an entry $A'[i,k]$ \emph{row-heavy} if the element $x \in X_i$ with $x = A'[i,k] \pm \Delta$ covers more than a $q/r$ fraction of the entries of the $i$-th row, that is, $\abs{\set{k' : A'[i,k'] = x \pm\Delta}} > nq/r$. 
            Similarly, we call an entry \emph{column-heavy} if its center $y\in Y_k$ satisfies $\abs{\set{i' : A'[i', k] = y \pm\Delta}} > nq/r$.
            Otherwise, we call $A'[i,k]$ \emph{light}.
            Let 
            \begin{equation*}
                A' = A'_\text{row} \sqcup A'_\text{col} \sqcup A'_\text{light}
            \end{equation*}
            denote the partition of the entries of $A'$ into row-heavy, column-heavy, and light entries. Entries that are both row-heavy and column-heavy are assigned arbitrarily to either part.
            Do the same for $B'$ and~$C'$.
            We proceed in two substeps.

            \begin{enumerate}
                \item \emph{(Light Entries)}
                    Each row or column bucket contains at most $nq/r$ light entries. An interval of width $2\Delta$ intersects at most two consecutive cover intervals, since their centers are more than~$2\Delta$ apart. In each row, split the light entries of every bucket into $O(q)$ groups of size at most~$n/(2r)$. Do the same independently for columns, and partition \smash{$A'_\text{light}$} according to the pairs of groups. This gives $O(q^2)$ matrices that are $(1/r,\Delta)$-regular.
                    Hence, we can transform the subinstance \smash{$(A'_\text{light}, B'_\text{light}, C'_\text{light})$} into $O(q^6)$ instances that are $(r,\Delta)$-uniform and $(1/r, \Delta)$-regular in a potential-adjusting way.
                    We insert all of these instances into $\mathcal I_\text{reg}$.

                    This step clearly runs in time $\tilde O(\abs{\mathcal I_\text{unif}}\cdot q^6n^2)$.

                \item \emph{(Heavy Entries)}
                    Focus on the row-heavy entries of $A'$ for now.
                    For each such entry $A'[i,k]$, we will update $A_\text{sm}[i,k] \gets A[i,k]$ and now also update its rank decomposition. Let \smash{$X_i^{\mathrm{heavy}} \subseteq X_i$} consist of the centers whose buckets contain more than $nq/r$ entries of the $i$-th row of $A'$. Since these buckets are disjoint, \smash{$|X_i^{\mathrm{heavy}}|\le r/q$}. This set $\Delta$-covers the $i$-th row of $A'_\text{row}$.
                    Therefore, we can trivially compute a $\Delta$-approximate rank-$(r/q)$ decomposition $(U', V', S')$ of $A'_\text{row}$.
                    We adjust this decomposition by
                    \begin{align*}
                        U[i,\ell] &\coloneqq U'[i,\ell] - u[i],\\
                        V[\ell,k] &\coloneqq V'[\ell,k] - v[k],
                    \end{align*}
                    which leads to a $\Delta$-approximate rank decomposition $(U,V,S = S')$ of the same rank of the restriction of $A$ to the row-heavy entries of $A'$. Indeed, for each row-heavy entry $(i,k)$, we have
                    \begin{align*}
                        A[i,k] &= A'[i,k] - u[i] - v[k] \\
                        &= U'[i,S'[i,k]] + V'[S'[i,k],k] - u[i] - v[k] \pm\Delta \\
                        &= U[i,S[i,k]] + V[S[i,k],k] \pm\Delta.
                    \end{align*}

                    We perform the same update-procedure for the column-heavy entries to update $A_\text{sm}$.
                    We do the same symmetrically for $B_\text{sm}$ and $C_\text{sm}$.
                    This step runs in time $\tilde O(\abs{\mathcal I_\text{unif}} \cdot n^2)$.
            \end{enumerate}

        \item \emph{(Recursion)}
            Finally, we recurse on the three instances  $(A_\text{sm}, B, C)$, $(A, B_\text{sm}, C)$, and $(A,B,C_\text{sm})$, which is possible since we have maintained approximate rank decompositions of the three matrices $A_\text{sm}$, $B_\text{sm}$, and $C_\text{sm}$.
            Denote the recursive outputs by $(\mathcal I_A, \mathcal T_A)$, $(\mathcal I_B, \mathcal T_B)$, and $(\mathcal I_C, \mathcal T_C)$.
            We return $(\mathcal I, \mathcal T) \coloneqq (\mathcal I_A \cup \mathcal I_B \cup \mathcal I_C \cup \mathcal I_\text{reg}, \mathcal T_A \cup \mathcal T_B \cup \mathcal T_C \cup \mathcal T_\text{sl} \cup \mathcal T_\text{unif})$. 
    \end{enumerate}
        This completes the algorithm description.
        Each step runs in the same time bound as the corresponding step in~\cite[Lemma~4.10]{Fischer26}. Therefore, the recursion analysis as well as the rank of subinstances is identical to \cite{Fischer26}. The same is true for the choice of parameters $t_\text{sl}, t_\text{unif}, q$. Thus, we omit further details here.
\end{proof}

\subsection{Uniform Regular to Uniform Small-Doubling}

\begin{restatable}[BSG Covering for Approximate Reals]{theorem}{bsgapprox}\label{thm:bsg_approx}
    Let $\Delta \ge 0$. Let $X,Y \subseteq \mathbb{R}$ be $\Omega(\Delta)$-spaced sets of size $n$ each, let $Z \subseteq \Real$ have size at most $n$, and let $L \ge 1$. Then, there are subsets $X_1, \ldots, X_L \subseteq X$ and $Y_1, \ldots, Y_L \subseteq Y$ such that
    \begin{enumerate}[label=(\roman*)]
        \item $\cabs{X_\ell + Y_\ell} \le O(L^5n)$ for all $\ell \in [L]$,
        \item $\abs{R} \le O(n^2/L)$ where $R \coloneqq \{(x,y) \in X \times Y : \exists z \in Z: x + y = z \pm \Delta\} \setminus \bigcup_{\ell = 1}^L (X_\ell \times Y_\ell)$.
    \end{enumerate}
    The sets $X_1, \ldots, X_L$, $Y_1, \ldots, Y_L$, and $R$ can be computed in randomized time $\tilde O(n^2)$.
\end{restatable}

We defer the proof of \Cref{thm:bsg_approx} to \Cref{sec:bsg}.

\newcommand{\round}[1]{\lfloor {#1}\rfloor_{\Delta}}
In this section, we will need some more lemmas corresponding to statements from additive combinatorics but for $\Delta$-approximate reals.
Therefore, let us introduce some notation and preliminary statements first to transfer the exact statements easily to the approximate setting.
When $\Delta \in \Realnneg$ is clear from context, define for any set $A\subseteq \Real$ the set $\round A \coloneqq \set{\Delta \cdot \lfloor a / \Delta \rfloor : a \in A}$, that is, the set of numbers in $A$ rounded down to an integer multiple of $\Delta$. (For $\Delta = 0$, define $\round A \coloneqq A$.)

\begin{lemma}\label{lem:rounded-size}
    Let $\Delta \in \Realnneg$ and $A \subseteq \Real$. Then, we have
    \begin{equation*}
        \cabs{A} \le \abs{\round A} \le 3\cabs{A}.
    \end{equation*}
\end{lemma}
\begin{proof}
    Notice that any pair of numbers $x, y \in \Real$ that are rounded to the same number $\Delta \cdot \lfloor x/\Delta\rfloor = \Delta \cdot \lfloor y/\Delta\rfloor$ satisfy $x = y \pm \Delta$. Thus, the rounded values themselves form a $\Delta$-cover of $A$.
    On the other hand, for any~$x', y' \in \Real$ satisfying $x' = y' \pm \Delta$ the rounded numbers $\lfloor x'/\Delta\rfloor$ and $\lfloor y'/\Delta\rfloor$ differ by at most 1. Consequently, each interval in a $\Delta$-cover of $A$ contains points from at most three rounding bins. This shows for any set $A \subseteq \Real$ that 
    \begin{equation*}
        \cabs{A} \le \abs{\round A} \le 3\cabs{A}. \qedhere
    \end{equation*}
\end{proof}

\begin{lemma}\label{lem:round-sum-commute}
    Let $\Delta \in \Realnneg$ and $A,B \subseteq \Real$. Then, we have
    \begin{equation*}
        \abs{\round A + \round B} \le 4\abs{\round{A+B}} \quad \text{and} \quad \abs{\round{A+B}} \le 4\abs{\round A + \round B}.
    \end{equation*}
\end{lemma}
\begin{proof}
    Let $a \in A$ and $b \in B$. Simple calculations yield
    \begin{equation*}
        \floor*{\frac{a}{\Delta}} + \floor*{\frac{b}{\Delta}} \le \frac{a}{\Delta} + \frac{b}{\Delta} = \frac{a+b}{\Delta} \le  \floor*{\frac{a+b}{\Delta}} + 1
    \end{equation*}
    and 
    \begin{equation*}
        \floor*{\frac{a + b}{\Delta}} \le \frac{a + b}{\Delta} = \frac{a}{\Delta} + \frac{b}{\Delta} \le \floor*{\frac{a}{\Delta}} + \floor*{\frac{b}{\Delta}} + 2.
    \end{equation*}
    This shows that the number in $\round A + \round B$ corresponds to at most 4 numbers in $\round{A+ B}$ and vice versa.
    Hence, we get
    \begin{equation*}
        \abs{\round A + \round B} \le 4\abs{\round{A+B}} \quad \text{and} \quad \abs{\round{A+B}} \le 4\abs{\round A + \round B}.\qedhere
    \end{equation*}
\end{proof}

\begin{lemma}[Ruzsa's Triangle Inequality for Approximate Reals]\label{lem:ruzsa-apx}
    Let $\Delta \in \Realnneg$. All finite sets $X,Y,Z \subseteq \Real$ satisfy
    \begin{equation*}
        \cabs{X+Y}\cabs{Z} \le 576 \cdot \cabs{X+Z}\cabs{Y+Z}.
    \end{equation*}
\end{lemma}
\begin{proof}
    The proof is a simple corollary of the exact version of Ruzsa's Triangle Inequality~\cite{Ruzsa96}, essentially the same statement with $\cabs{\cdot}$ replaced by $\abs{\cdot}$ and without an additional constant. Assume $\cabs{Z} > 0$. We have
    \begin{align*}
        \cabs{X+Y} &\le \abs{\round{X+Y}}\\
        &\le 4\cdot\abs{\round X + \round Y} \\
        &\overset{(\star)}{\le} 4\cdot\frac{\abs{\round X +\round Z} \cdot \abs{\round Y + \round Z}}{\abs{\round Z}} \\
        &\le 64\cdot\frac{\abs{\round{X+Z}}\cdot \abs{\round{Y+Z}}}{\abs{\round Z}} \\
        &\le 576\cdot \frac{\cabs{X+Z}\cdot \cabs{Y+Z}}{\cabs{Z}},
    \end{align*}
    where we used Ruzsa's Triangle Inequality for the step ($\star$). All other steps are applications of \Cref{lem:rounded-size,lem:round-sum-commute}.
\end{proof}

Finally, we prove the goal of this section, the reduction from approximately uniform and regular instances to approximately uniform and low-doubling instances.
We say that a $(D, \Delta)$-uniform instance $(A,B,C)$ is \emph{$(K, \Delta)$-doubling} if the set $X$ of edge weights in $A$, $B$, and $C$ satisfies that $\cabs{X + X} \leq K \cabs{X}$.

\begin{lemma}[Reduction from Approx.\ Uniform Regular to Approx.\ Uniform Low-Doubling Instances]\label{lem:uniform-regular-to-small-doubling}
    Let $\Delta \in \Realnneg$ and $K \ge 1$. There is a potential-adjusting reduction from a Real Exact Triangle Instance $(A,B,C)$ that is $(D, \Delta)$-uniform and $(1/D,\Delta)$-regular to $O(K^{1/56})$ instances that are $(D,\Delta)$-uniform and $(K, \Delta)$-doubling. It runs in time \[\tilde O\parens*{n^2K^{1/56} + \frac{n^3}{K^{1/56}}}.\]
\end{lemma}
\begin{proof}
    Denote by $X$ the greedy minimum $\Delta$-cover of the set of entries of the instance; its centers are more than $2\Delta$ apart. By uniformity, we know that $\abs{X} \le D$.
    We apply BSG Covering (\Cref{thm:bsg_approx}) with input sets $X,X,-X$, tolerance $4\Delta$, size parameter $|X|$, and some parameter $L$ to be determined later.
    This results in sets $X_1, \ldots, X_L, Y_1, \ldots, Y_L$ and $R$. Since each interval of width $8\Delta$ is covered by four intervals of width $2\Delta$, the returned sets satisfy $|X_\ell+Y_\ell|_{\pm\Delta}\le4|X_\ell+Y_\ell|_{\pm4\Delta}\le O(L^5D)$ and $|R|\le O(D^2/L)$. For each $\Delta$-approximate exact triangle with representatives $x,y,z\in X$, we have $x+y=-z\pm4\Delta$. Therefore, its representative pair $(x, y)$ is either in the \emph{structured part} (if $(x, y) \in X_\ell \times Y_\ell$ for some $\ell$) or in the \emph{remainder} if $(x, y) \in R$. We deal with both types of triangles separately.

    \begin{enumerate}
        \item \emph{(Remainder)}
            We explicitly list all $\Delta$-approximate exact triangles in the remainder.
            Enumerate all pairs $(x,y) \in R$ and enumerate all $k \in [n]$.
            Then, enumerate all $i \in [n]$ with $A[i,k] = x \pm \Delta$ and all $j \in [n]$ with $B[k,j] = y \pm \Delta$.
            We store all triangles enumerated this way in $\mathcal T$.

            Clearly, we enumerate all triangles in the remainder. For the running time, recall that the instance is $(1/D, \Delta)$-regular, hence for a fixed $k$ and $x$, there are at most $n/D$ valid choices of $i$, and similarly for $y$ and $j$. Therefore, the running time is bounded by
            \begin{equation*}
                \tilde O\parens*{\abs{R}\cdot n \cdot \frac{n}{D}\cdot \frac{n}{D}} = \tilde O\parens*{\frac{n^3}{L}}.
            \end{equation*}
            
        \item \emph{(Structured Part)}
            Enumerate all $\ell \in [L]$. First discard from $X_\ell$ every center whose interval of width~$2\Delta$ contains no entry of $A$, and do the same for $Y_\ell$ and $B$. We compute the set $X_\ell + Y_\ell$ by brute-force and construct the matrices
            \begin{align*}
                A_\ell[i,k] &\coloneqq \begin{cases} \mathmakebox[1cm]{A[i,k]} &\text{if $A[i,k] = x \pm \Delta$ for some $x \in X_\ell$,}\\\bot&\text{otherwise;}\end{cases}\\
                B_\ell[k,j] &\coloneqq \begin{cases} \mathmakebox[1cm]{B[k,j]} &\text{if $B[k,j] = y \pm \Delta$ for some $y \in Y_\ell$,}\\\bot&\text{otherwise;}\end{cases}\\
                C_\ell[i,j] &\coloneqq \begin{cases} \mathmakebox[1cm]{C[i,j]} &\text{if $C[i,j] = -z \pm 3\Delta$ for some $z \in X_\ell + Y_\ell$,}\\\bot&\text{otherwise.}\end{cases}
            \end{align*}

            This way, each $\Delta$-approximate exact triangle in the structured part appears in some instance $(A_\ell, B_\ell, C_\ell)$. 
            We further distinguish two cases.

            \begin{enumerate}
                \item \emph{(Sparse Case: $\abs{X_\ell} \le D/L^2$ or $\abs{Y_\ell} \le D/L^2$)}
                    In this case, one of the matrices $A_\ell$ or $B_\ell$ is sparse. Suppose $\abs{X_\ell} \le D/L^2$. Then, all entries in $A_\ell$ are covered by a $\Delta$-cover of size at most $D/L^2$ and each element in the cover covers at most a $1/D$-fraction of the entries, by regularity. Hence, $A_\ell$ can contain at most $n^2/L^2$ non-$\bot$ entries.
                    We can therefore solve the instance $(A_\ell, B_\ell, C_\ell)$ by brute-force and place all triangles in $\mathcal T$ in time $O(n^3/L)$ across all iterations.
                
                \item \emph{(Dense Case: $\abs{X_\ell} > D/L^2$ and $\abs{Y_\ell} > D/L^2$)}
                    Let $W_\ell$ be the set of weights in $(A_\ell,B_\ell,C_\ell)$ and let $H_\ell\coloneqq X_\ell\cup Y_\ell$. The set
                    \[
                        Z_\ell \coloneqq \parens*{ H_\ell \cup (-(X_\ell+Y_\ell)) } + \set{-3\Delta, -2\Delta, \ldots, 3\Delta}
                    \]
                    is a $\Delta$-cover of $W_\ell$. Moreover, every center in $H_\ell$ is within $\Delta$ of a weight in $W_\ell$, by the removal of unused centers. Expanding each interval in a minimum $\Delta$-cover of $W_\ell$ to width $4\Delta$ therefore covers $H_\ell$. Each expanded interval contains at most two centers, since these are more than $2\Delta$ apart. Thus $\cabs{H_\ell}=|H_\ell|\le2\cabs{W_\ell}$. We will bound $\cabs{Z_\ell+Z_\ell}$ relative to $\cabs{H_\ell}$.

                    First, note that we can bound the approximate doubling of $X_\ell$ using Ruzsa's triangle inequality for approximate reals (\Cref{lem:ruzsa-apx}) by
                    \begin{equation*}
                        \cabs{X_\ell + X_\ell} \le 576\cdot \frac{\cabs{X_\ell + Y_\ell}\cabs{X_\ell + Y_\ell}}{\cabs{Y_\ell}} \le O\parens*{\frac{L^{10}D^2}{D/L^2}} = O(L^{12}D) \le O(L^{14})\cabs{X_\ell}.
                    \end{equation*}
                    Symmetrically, $\cabs{Y_\ell+Y_\ell}\le O(L^{12}D)$. Since $H_\ell+H_\ell=(X_\ell+X_\ell)\cup(X_\ell+Y_\ell)\cup(Y_\ell+Y_\ell)$, it follows that
                    \[
                        \cabs{H_\ell+H_\ell}\le O(L^{12}D)\le O(L^{14})\cabs{H_\ell}.
                    \]
                    Let $Q_\ell\coloneqq\round{H_\ell}$. By \Cref{lem:rounded-size,lem:round-sum-commute}, we have $|Q_\ell+Q_\ell|\le O(L^{14})|Q_\ell|$. The Pl\"unnecke--Ruzsa inequality gives
                    \[
                        |aQ_\ell-bQ_\ell|\le O(L^{14(a+b)})|Q_\ell|,
                    \]
                    where $a,b$ are nonnegative integers and $aQ_\ell$ denotes the $a$-fold sumset. Every element of $Z_\ell+Z_\ell$ lies within $O(\Delta)$ of an element of $(Q_\ell+Q_\ell)\cup(Q_\ell-2Q_\ell)\cup(-4Q_\ell)$, as each term from $H_\ell$ contributes less than $\Delta$ of rounding error. Consequently, $\cabs{Z_\ell+Z_\ell}\le O(L^{56})|Q_\ell|\le O(L^{56})\cabs{H_\ell}$. Since~$Z_\ell$ $\Delta$-covers $W_\ell$, we obtain
                    \begin{equation*}
                        \cabs{W_\ell+W_\ell}\le3\cabs{Z_\ell+Z_\ell}\le O(L^{56})\cabs{W_\ell}.
                    \end{equation*}
                    We pick $L = \Theta(K^{1/56})$ such that $W_\ell$ has approximate doubling at most $K$.
                    Therefore, the whole instance $(A_\ell, B_\ell, C_\ell)$ is $(K, \Delta)$-doubling and $(D,\Delta)$-uniform, as desired, and we place it in $\mathcal I$.

                    The arguments in this step are just existential, so the total running time across all iterations is bounded by $O(n^2L)$.
            \end{enumerate}
    \end{enumerate}

    This completes the description of the reduction, which is clearly potential-adjusting with $\abs{\mathcal I} \le O(K^{1/56})$.
    For the running time, the initial BSG Covering takes time $O(D^2) \le O(n^2)$.
    The remaining steps are analyzed before.
    In total, the running time is bounded by
    \begin{equation*}
        \tilde O\parens*{n^2L + \frac{n^3}{L}} = \tilde O\parens*{n^2K^{1/56} + \frac{n^3}{K^{1/56}}}.\qedhere
    \end{equation*}
\end{proof}

\subsection{Putting the Pieces Together}

\weightreduction*
\begin{proof}[Proof of \Cref{lem:low-rank-to-small-doubling}]
    First, apply \Cref{lem:low-rank-to-uniform-regular} with $t \coloneqq s^{1/100000}$ to get $t^{875}n^{o(1)}$ instances. 
    For each resulting instance, run \Cref{lem:uniform-regular-to-small-doubling} with parameter $K \coloneqq s$. This results in a total of $t^{875}n^{o(1)}\cdot s^{1/56} \le sn^{o(1)}$ instances that are $(r,\Delta)$-uniform and $(s,\Delta)$-doubling, as claimed. 
    The total running time is
    \begin{equation*}
        \parens*{n^2rt^{875} + \frac{n^3}{t} + t^{875}\cdot \parens*{n^2s^{1/56} + \frac{n^3}{s^{1/56}}}}^{1+o(1)} = \parens*{n^2 r s + \frac{n^3}{s^{1/100000}}}^{1+o(1)}.\qedhere
    \end{equation*}
\end{proof}

\subsection{Proof of the BSG Theorem for Approximate Reals}
\label{sec:bsg}

We follow the proof and exposition by Chan and Lewenstein~\cite{ChanL15}.
The proof is based on a graph lemma, which Chan and Lewenstein prove in their paper. For the purpose of our proof, we can simply reuse it without any difference. As they do, we prove the BSG Theorem in the single set version first and then use that to prove \Cref{thm:bsg_approx}.

\begin{lemma}[Graph Lemma~\protect{\cite[Lemma~7.1]{ChanL15}}]\label{lem:graph_lemma}
    Let $\alpha > 0$ and let $X, Y \subseteq \Real$ be sets of size $n$ each.
    Given a bipartite graph $G \subseteq X \times Y$, with $|G| \ge \alpha n^2$, there exist $X' \subseteq X$ and $Y' \subseteq Y$ such that
    \begin{enumerate}[label=(\roman*)]
         \item for every $x' \in X'$, $y' \in Y'$, there are $\Omega(\alpha^5n^2)$ length-3 paths from $x'$ to $y'$, and
         \item $|G \cap (X' \times Y')| \ge \Omega(\alpha|X'|n) \ge \Omega(\alpha^2n^2)$.
     \end{enumerate}
     The sets can be computed in randomized time $\tilde O((1/\alpha)^5|X'| + (1/\alpha)n + (1/\alpha)^6)$.
\end{lemma}

\begin{lemma}[Single-Set BSG for Approximate Reals]\label{thm:bsg_intervals}
    Let $\Delta \in \Realnneg$ and let $X, Y \subseteq \mathbb{R}$ be $\Omega(\Delta)$-spaced sets of size $n$ each. Let $G \subseteq X \times Y$ be of size $\abs{G} \ge \alpha n^2$ and suppose that $\cabs{\set{x + y : (x,y) \in G}} \le n$.
    
    Then, there are $X' \subseteq X$ and $Y' \subseteq Y$ such that
    \begin{enumerate}[label=(\roman*)]
        \item $\cabs{X' + Y'} \le O((1/\alpha)^5 n)$ and
        \item $\abs{G \cap (X' \times Y')} \ge \Omega(\alpha^2n^2)$.
    \end{enumerate}
\end{lemma}
\begin{proof}
    Fix a constant $c_0\ge1$ such that $X$ and $Y$ are $\Delta/c_0$-spaced. Consider the bipartite graph with $X$ and $Y$ as its two sides and edges given by $G$.
    Let $X'$ and $Y'$ be the sets given by the graph lemma (\Cref{lem:graph_lemma}).
    We claim that these sets satisfy both claimed properties.
    Property (ii) immediately follows from \Cref{lem:graph_lemma}, property (ii), so we focus on proving the first property.

    Let $S$ be a $\Delta$-cover of the set $\set{x + y : (x,y) \in G}$ of size $n$.
    Let $T$ be the greedy minimum $\Delta$-cover of $X' + Y'$, whose centers are more than $2\Delta$ apart.
    We analyze the following algorithm to bound the size of $T$:

    \begin{algorithm}[H]
    \DontPrintSemicolon
    \SetKwFor{ForEach}{for each}{do}{end for}
    \SetArgSty{textnormal}
    
    \ForEach{$t \in T$}{
      take the lexicographically smallest $(x',y')\in X'\times Y'$ with $x' + y' = t \pm \Delta$\;
      \ForEach{length-3 path $x', y, x, y'$ in $G$}{
        mark every triple $(a,b,c) \in S^3$ with
        \begin{enumerate}[itemsep=0pt, parsep=0pt, topsep=0pt]
            \item $x' + y  = a \pm\Delta$,
            \item $x + y  = b \pm \Delta$, and
            \item $x + y'  = c \pm \Delta$
        \end{enumerate}
      }
    }
    \end{algorithm}

    By property (i) of \Cref{lem:graph_lemma}, the innermost step is executed at least $\Omega(|T| \cdot \alpha^5|X||Y|)$ times. Since $(x',y)$, $(x,y)$, and $(x,y')$ are all edges in $G$ and by the definition of $S$, at least one triple is marked in each iteration.
    Next, we argue that each element in $S^3$ is marked $O(c_0^2)=O(1)$ times, bounding the number of total marks by $O(|S|^3)$. This then immediately implies \[|T| \le O\parens*{\frac{c_0^2|S|^3}{\alpha^5 |X||Y|}} = O\parens*{\frac{n}{\alpha^5}},\] and since $T$ is a $\Delta$-cover of $X' + Y'$, we have $\cabs{X' + Y'} \le O((1/\alpha)^5n)$, property (i).

    To prove the claim, consider $(a,b,c) \in S^3$ that is marked by some path of length 3 via $x', y, x, y'$. 
    By adding the conditions on $a$ and $c$ and subtracting the condition on $b$, we get
    \begin{equation*}
        a - b + c = x' + y' \pm 3\Delta,
    \end{equation*}
    and $a-b+c = t \pm 4\Delta$.
    Hence, there can be at most $O(1)$ choices for $t \in T$ leading to $(a,b,c)$ being marked.
    Each choice of $t$ in turn uniquely determines $(x', y') \in X' \times Y'$.
    Furthermore, we know that $x = c - y' \pm 2\Delta$ and $y = a - x' \pm 2\Delta$.
    The $\Delta/c_0$ spacing gives $O(c_0)$ options each for $x$ and $y$ for a fixed choice of $(x',y')$. Thus $(a,b,c)$ can be marked at most $O(c_0^2)=O(1)$ times.
\end{proof}

\bsgapprox*
\begin{proof}[Proof of \Cref{thm:bsg_approx}]
    If $n \le L^5$, simply return $X_1 \coloneqq X$, $Y_1 \coloneqq Y$, and leave all other sets empty.
    We repeatedly apply \Cref{thm:bsg_intervals} to extract the sets as follows.

    We initialize the set $G \subseteq X \times Y$ as
    \begin{equation*}
        G \coloneqq \set{(x,y) \in X \times Y : \text{there exists $z \in Z$ with $x + y = z \pm \Delta$}}.
    \end{equation*}
    Then we run iterations $i = 1, 2, \ldots, L$, where we fix the sets $X_i$ and $Y_i$ in iteration $i$. Initially, they are all empty.
    
    In an iteration $i$, we first check if $n^2/\abs{G} \ge c\cdot L$ for some constant $c$ determined later and in this case simply return with $R \coloneqq G$.
    Otherwise, we apply \Cref{thm:bsg_intervals} on $X$ and $Y$ with the current graph $G$ and $\alpha = \abs{G}/n^2$ to receive sets $X_i$ and $Y_i$.
    Afterwards, we remove $(X_i \times Y_i)$ from $G$ and continue with the next iteration. 

    \paragraph{Correctness.}
    Notice that the set of sums represented by $G$ has a $\Delta$-cover of size at most $n$, namely $Z$, which is still valid even after removing elements from $G$. Hence we can apply \Cref{thm:bsg_intervals}. The bounds on $\cabs{X_\ell + Y_\ell}$ and $\abs{R}$ are clearly satisfied.
    All that remains to show is that the algorithm terminates after $L$ iterations.
    Let $G_i$ denote the set $G$ in iteration $i$.
    Using \Cref{thm:bsg_intervals}, property (ii), we can bound its size by 
    \begin{equation*}
        \abs{G_{i+1}} \le \abs{G_i} - \Omega\parens*{\parens*{\frac{\abs{G_i}}{n^2}}^2n^2} = \abs{G_i} \cdot \parens*{1 - \Omega\parens*{\frac{\abs{G_i}}{n^2}}}.
    \end{equation*}
    Following the analysis in~\cite{ChanL15}, this implies
    \begin{equation*}
        \frac{n^2}{\abs{G_{i+1}}} \ge \frac{n^2}{\abs{G_i}} \cdot \parens*{1 + \Omega\parens*{\frac{\abs{G_i}}{n^2}}} \ge \frac{n^2}{\abs{G_i}} + \Omega(1).
    \end{equation*}
    By iterating, we get $n^2/\abs{G_L} \ge \Omega(L)$ and can hence choose a $c$ to terminate in the correct number of iterations.
    
    The running time analysis is analogous to the analysis in~\cite{ChanL15}.
\end{proof}

\bibliographystyle{alphaurl}
\bibliography{main}

\end{document}